\documentclass[12pt,a4paper]{article}
\usepackage[T1]{fontenc}
\usepackage{amsmath,amsfonts,amssymb,amsthm,mathabx}
\usepackage[latin1]{inputenc}
\usepackage{cite}
\usepackage{hyperref}
\usepackage{refstyle}
\usepackage{dsfont}
\usepackage{graphicx}
\usepackage{color}
\usepackage{stmaryrd}
\usepackage[multiple]{footmisc}
\usepackage{here}
\usepackage{tgtermes} 
\usepackage{tgpagella} 
\usepackage{sfmath}
\usepackage[british]{babel}

\newcommand{\Ad}{\mathrm{Ad}}
\newcommand{\ad}{\mathrm{ad}}

\def\tr{{\rm Tr}}
\renewcommand{\d}{\partial}

\newcommand{\Diff}{\mathrm{Diff}^+(S^1)}
\newcommand{\hatDiff}{\widehat{\mathrm{Diff}^+}(S^1)}

\def\Vect{\text{\fontfamily{pzc}\selectfont vect}\left(S^1\right)}
\def\hatVect{\widehat{\text{\fontfamily{pzc}\selectfont vect}}\left(S^1\right)}
\newcommand{\BMS}{\mathrm{BMS}_3}
\newcommand{\hatBMS}{\widehat{\mathrm{BMS}}_3}

\def\bms{\text{\fontfamily{pzc}\selectfont bms}_3}
\def\hatbms{\widehat{\text{\fontfamily{pzc}\selectfont bms}}_3}
\def\bmsalg{\text{\fontfamily{pzc}\selectfont bms}}

\newcommand{\MBMS}{{\rm Maxwell}\textrm{-}\mathrm{BMS}_3}
\newcommand{\MhatBMS}{{\rm Maxwell}\textrm{-}\widehat{\mathrm{BMS}}_3}

\def\mbms{ \text{\fontfamily{pzc}\selectfont maxwell} \textrm{-}  \text{\fontfamily{pzc}\selectfont bms}_3}
\def\mhatbms{ \text{\fontfamily{pzc}\selectfont maxwell} \textrm{-}  \widehat{\text{\fontfamily{pzc}\selectfont bms}}_3}
\def\Sl{  {\rm{SL}}\left(2,\mathbb{R}\right)  } 

\def\sl{  \mathfrak{sl}\left(2,\mathbb{R}\right)  } 
\def\sl{  \mathfrak{sl}\left(2,\mathbb{R}\right)  } 
\def\slab{  \mathfrak{sl}\left(2,\mathbb{R}\right)^{{\rm(ab)}}  } 
\def\slabe{  \mathfrak{sl}\left(2,\mathbb{R}\right)^{{\rm(ab)}}_{{\rm ext}}  }

\def\ISl{  {\rm{ISL}}\left(2,\mathbb{R}\right)  }

\newcommand{\Maxd}{\mathrm{Maxwell}_D}
\newcommand{\Max}{\mathrm{Maxwell}_3}
\newcommand{\dMax}{\overline{\mathrm{Maxwell}}_3}

\def\maxd{\text{\fontfamily{pzc}\selectfont maxwell}_D}
\def\max{\text{\fontfamily{pzc}\selectfont maxwell}_3}

\def\LSl{  L{\rm{SL}}\left(2,\mathbb{R}\right)  } 
\def\hatLSl{  \widehat{L\,{\rm{SL}}}\left(2,\mathbb{R}\right)  } 

\def\Lsl{  L\,{\mathfrak{sl}}\left(2,\mathbb{R}\right)  } 
\def\hatLsl{ \widehat{L\, \mathfrak{sl}} \left(2,\mathbb{R}\right)  }

\def\LISl{  L{\rm{ISL}}\left(2,\mathbb{R}\right)  } 
\def\hatLISl{  \widehat{L\,{\rm{ISL}}}\left(2,\mathbb{R}\right)  }

\newcommand{\LMax}{L\mathrm{Maxwell}_3}
\newcommand{\hatLMax}{\widehat{L\mathrm{Maxwell}}_3}
\def\Lmax{L \text{\fontfamily{pzc}\selectfont maxwell}_3}
\def\hatLmax{\widehat{L \text{\fontfamily{pzc}\selectfont maxwell}}_3}

\def\so{  \mathfrak{so} } 
\def\slt{  \mathfrak{sl}\left(3,\mathbb{R}\right)  } 
\def\sln{  \mathfrak{sl}\left(N,\mathbb{R}\right)  } 

\newcommand{\mfg}{\mathfrak{g}}
\newcommand{\mfk}{\mathfrak{k}}
\newcommand{\mfh}{\mathfrak{h}}
\newcommand{\mfm}{\mathfrak{m}}
\def\cv{\text{\fontfamily{pzc}\selectfont V\,}}
\def\cvt{\tilde{\cv}}

\def\a{\alpha}
\def\b{\beta}

\def\cC{\mathcal{C}}

\def\cJ{\mathcal{J}}
\def\cS{\mathcal{S}}

\newcommand{\mfs}{\mathfrak{s}}

\def\bJ{\boldsymbol{J}}
\def\bPi{\boldsymbol{\Pi}}
\def\bF{\boldsymbol{F}}
\def\bT{\boldsymbol{T}}
\def\bk{\boldsymbol{k}}
\def\bc{\boldsymbol{c}}
\def\bTT{\boldsymbol{\mathcal{T}}}
\def\bt{\boldsymbol{t}}
\def\bell{\boldsymbol{\ell}}
\def\bJJ{\boldsymbol{\mathcal{J}}}
\def\bPP{\boldsymbol{\mathcal{P}}}
\def\bFF{\boldsymbol{\mathcal{F}}}

\def\be{\begin{equation}}
\def\ee{\end{equation}}
\def\ba{\begin{array}{lcl}}
\def\ea{\end{array}}

\advance\textheight\topskip
\numberwithin{equation}{section}

\begin{document}

\title{\sc \LARGE{\textbf{The Maxwell group in 2+1 dimensions\\and its infinite-dimensional enhancements}}\vspace{-1cm}}

\date{}
\maketitle
%\vspace{5cm}
%\def\mytitle{\sc \textbf{A note on Maxwell groups in 2+1 dimensions\\and their infinite-dimensional extensions}}
%\pagestyle{myheadings} \markboth{\textsc{\small  P.~Salgado-Rebolledo}}{%
 %\textsc{\small Title}}
\addtolength{\headsep}{4pt}

\begin{centering}

%\vspace{1cm}
%\textbf{\Large{\mytitle}}
%\vspace{1.5cm}

 {\large Patricio Salgado-Rebolledo }\\
 \vspace{.5cm}
\small{\tt patricio.salgado@pucv.cl}

\vspace{.5cm}

\begin{minipage}{.9\textwidth}\small \it  \begin{center}
   Instituto de F\'isica, Pontificia Universidad Cat\'olica de Valpara\'iso\\ Casilla 4059, Valpara\'iso, Chile 
 \end{center}
\end{minipage}

\end{centering}

\vspace{2cm}

\begin{center}
\begin{minipage}{.9\textwidth}
\textsc{\textbf{Abstract}}. The Maxwell group in 2+1 dimensions is given by a particular extension of a semi-direct product. This mathematical structure provides a sound framework to study different generalizations of the Maxwell symmetry in three space-time dimensions. By giving a general definition of extended semi-direct products, we construct infinite-dimensional enhancements of the Maxwell group that enlarge the ${\rm ISL}(2,\mathbb{R})$ Kac-Moody group and the $\hatBMS$ group by including non-commutative supertranslations. The coadjoint representation in each case is defined, and the corresponding geometric actions on coadjoint orbits are presented. These actions lead to novel Wess-Zumino terms that naturally realize the aforementioned infinite-dimensional symmetries. We briefly elaborate on potential applications in the contexts of three-dimensional gravity, higher-spin symmetries, and quantum Hall systems.
 \end{minipage}
\end{center}

\vspace{1cm}
\thispagestyle{empty}

\newpage

\begin{small}
{\addtolength{\parskip}{-2pt}
 \tableofcontents}
\end{small}
\thispagestyle{empty}

\newpage

\section{Introduction}

The Maxwell algebra is an extension of the Poincar\'e symmetry that naturally arises in the study of particle systems in the presence of a constant electromagnetic field. It includes, apart from translations and Lorentz transformations, an Abelian ideal spanned by a second rank antisymmetric Lorentz tensor $\bF_{\mu\nu}$, associated to the electromagnetic field strength $F_{\mu\nu}=\partial_\mu A_\nu -\partial_\nu A_\mu$. This enlargement stems from the fact that minimal coupling of a particle to the electromagnetic field requires to modify the canonical momentum in the form $p_\mu\rightarrow \pi_\mu = p_\mu +A_\mu$. When the electromagnetic field is constant (i.e. homogeneous and static), it can be thought of as an extension of Minkowski space, whose isometries include non-commutative translations. This leads to the Maxwell algebra:
\be\eqlabel{maxwellalg}
\ba
\left[\bJ_{\mu\nu},\bJ_{\rho\sigma}\right]=\eta_{\mu\rho}\bJ_{\nu\sigma}+\eta_{\nu\sigma}\bJ_{\mu\rho}-\eta_{\mu\sigma}\bJ_{\nu\rho}-\eta_{\nu\rho}\bJ_{\mu\sigma}\,, & \quad & \left[\bPi_{\text{\ensuremath{\mu}}},\bPi_{\nu}\right]= \bF_{\mu\nu}\,,\\
\left[\bJ_{\mu\nu},\bPi_{\rho}\right]=\eta_{\mu\rho}\bPi_{\nu}-\eta_{\nu\rho}\bPi_{\mu}\,, & \quad & \left[\bPi_{\mu},\bF_{\nu\rho}\right]=0\,,\\
\left[\bJ_{\mu\nu},\bF_{\rho\sigma}\right]=\eta_{\mu\rho}\bF_{\nu\sigma}+\eta_{\nu\sigma}\bF_{\mu\rho}-\eta_{\mu\sigma}\bF_{\nu\rho}-\eta_{\nu\rho}\bF_{\mu\sigma}\,, & \quad & \left[\bF_{\mu\nu},\bF_{\rho\text{\ensuremath{\sigma}}}\right]=0\,.
\ea
\ee
The Maxwell group is given by the semi-direct product of the Lorentz group and a centrally extended space of translations. It was first studied in \cite{Schrader:1972zd} as the covariance group of the (3+1)-dimensional Klein Gordon and Dirac equations associated to a particle moving in a constant electromagnetic field\footnote{For a detailed analysis including the non-relativistic case, see \cite{Beckers:1983gp}, and for a geometric derivation of wave equations for particles coupled to constant electromagnetic field based on local representations of the Maxwell group, see \cite{Hoogland:1978wi,negro1990local1}.}. However, previous indications of this symmetry were  found in \cite{Bacry:1970ye}, where the kinematic symmetry of this kind of particle systems was shown to include a central extension of the translation group. This extension was characterized by the electric charge in pretty much the same way as the mass appears in the central generator of the Bargmann algebra.

Later on, the Maxwell algebra was used in the context of gravity. By gauging it, one can define a modified gravitational theory that extends General Relativity by including a generalized cosmological term  \cite{deAzcarraga:2010sw}. In this case, the components of the antisymmetric tensor $\bF_{\mu\nu}$ are not related to an electromagnetic field, but reinterpreted as extra generators associated to the new field content of the theory. This provides a geometric framework to introduce vector inflatons in cosmological models \cite{deAzcarraga:2012qj} 

In the last years, different generalizations to higher dimensions, supersymmetric extensions, and deformations of the Maxwell algebra have been formulated\footnote{The supersymmetric extension of the Maxwell algebra has been used to construct supergravity models \cite{Penafiel:2017wfr,Bonanos:2010fw,Ravera:2018vra,Concha:2018ywv}, as well as in the description of the superparticle on constant backgrounds \cite{Bonanos:2009wy}.}\cite{Soroka:2004fj,Bonanos:2009wy,Gomis:2009dm,Gibbons:2009me,Khasanov:2011jr,Durka:2011va,Kamimura:2011mq,Fedoruk:2013sna}. By means of algebra expansion methods \cite{deAzcarraga:2007et,Izaurieta:2006zz}, different families of Lie algebras that further extend the commutation relations \eqref*{maxwellalg} have been constructed, leading to novel (super)gravity theories in diverse dimensions \cite{deAzcarraga:2012zv,Concha:2013uhq,Salgado:2014qqa,Gonzalez:2014tta,Concha:2014xfa,Concha:2014zsa,Durka:2016eun,Gonzalez:2016xwo}. Moreover, the Maxwell algebra can be understood as the simplest Poincar\'e extension within an infinite-dimensional free algebra, which describes the motion of a charge distribution in an generic electro-magnetic field through multipole expansion \cite{Bonanos:2008ez,Gomis:2017cmt}.

The (2+1)-dimensional version of the Maxwell symmetry naturally arises when studying anyons coupled to a constant external electromagnetic field \cite{Negro:2005jp,Brzykcy:2017esn}. This is of particular interest in the study of quantum Hall systems, where magnetic translations associated to an electric charge in presence of a constant magnetic field $B$ have non-vanishing Poisson brackets \cite{Duval:2000xr,Horvathy:2004am}
\begin{equation}
\left\{ \Pi_x , \Pi_y \right\}= B \, .
\end{equation}
In fact, in \cite{Palumbo:2016nku} it is shown that the Maxwell algebra properly describes the Girvin-MacDonald-Plazmann algebra of magnetic translations in quantum Hall systems \cite{girvin1985collective}. Furthermore, a relativistic version of the Wen-Zee term was constructed by considering a Chern-Simons action invariant under the Maxwell algebra as an effective theory for the boundary dynamics of a topological insulator.

On the other hand, Chern-Simons forms are interesting in the context of gravitational theories, as they provide a gauge formulation of three-dimensional Einstein gravity \cite{Witten:1988hc,Zanelli:2012px}. Along these lines, a Maxwell-invariant Chern-Simons action defines an extended theory for gravity in 2+1 dimensions\footnote{A Chern-Simons theory invariant under the Maxwell algebra was previously considered in \cite{Cangemi:1992ri,Duval:2008tr}, where it was used to obtain the Cangemi-Jackiw action for gravity in 1+1 dimensions \cite{Cangemi:1992bj,Cangemi:1993sd} by means of dimensional reduction.}\footnote{Supersymmetric extensions, higher-spin extensions, and non-relativistic limits of Chern-Simons gravity theories invariant under the Maxwell algebra have been constructed in \cite{Concha:2015woa,Caroca:2017izc,Aviles:2018jzw}, respectively.}\cite{Salgado:2014jka,Hoseinzadeh:2014bla,Concha:2018zeb}.
In \cite{Concha:2018zeb}, it was shown that, given suitable boundary conditions, the asymptotic symmetry of such extended theory is given by an infinite-dimensional enhancement of the Maxwell algebra in 2+1 dimensions. This algebra can be understood as an Maxwell-like extension of the $\hatbms$ algebra\footnote{A semi-simple enlargement of $\hatbms$, related with the above-mentioned Maxwell extension by an In{\"o}n{\"u}-Wigner contraction, was also studied in \cite{Concha:2018jjj}.}, known from the analysis of asymptotically flat three-dimensional Einstein gravity \cite{Barnich:2006av}. Moreover, the same extended $\hatbms$ symmetry was previously obtained as a particular expansion of the Virasoro algebra, and can also be recovered by means of a generalized Sugawara construction of a Maxwell-Kac-Moody algebra \cite{Caroca:2017onr}. The Maxwell-Kac-Moody algebra, in turn, is by definition the symmetry of the Wess-Zumino-Witten (WZW) model that follows from a Maxwell-invariant Chern-Simons action after Hamiltonian reduction. The fact that both infinite-dimensional symmetries can be obtained from a Maxwell-invariant Chern-Simons theory suggests that they could also be relevant in the context of quantum Hall systems. Indeed, the analogy between three-dimensional gravity and the quantum Hall effect has been already considered in the literature (see, for instance, \cite{Myung:1998jw,chakraborty2011taking}). As, so far, only the algebras are known, a natural question is: what is the group structure behind these infinite-dimensional symmetries?

In order to answer this question, it is important to note that both the Maxwell-Kac-Moody algebra and the Maxwell-like extension of the $\hatbms$ algebra found in \cite{Caroca:2017onr} extend their Poincar\'e-like partners in that they include non-commutative supertranslations. On the other hand, the $\ISl$ Kac-Moody symmetry as well as the $\hatBMS$ symmetry correspond to semi-direct product groups\cite{Barnich:2014kra,Barnich:2015uva,Barnich:2017jgw}. Therefore, the group structure behind the aforementioned Maxwell-like symmetries must be given by a general mathematical construction that extends semi-direct products. In this article we show that this is indeed the case.

In this paper, we study the group structure of the Maxwell symmetry in 2+1 dimensions and its infinite-dimensional generalizations. The construction is based on a non-central extension of a semi-direct product group. Following the general construction of semi-direct products \cite{rawnsley1975representations,Baguis:1997yp}, the adjoint and the coadjoint representations of the extension here presented are given. The Kirillov-Kostant-Souriau symplectic structure on coadjoint orbits is defined, and the Hamiltonian geometric action associated to an extended semi-direct product is computed along the lines of \cite{Alekseev:1988ce,Delius:1990pt,Alekseev:1993qs,Alekseev:2018ful} (for a recent discussion in the context of three-dimensional gravity see \cite{Barnich:2017jgw}). Based on these definitions, we derive the Maxwell group in $D$ dimensions as a particular case of extended semi-direct product and the action of a relativistic particle in a constant electromagnetic field as the corresponding geometric action. Subsequently, we focus on the (2+1)-dimensional case. Following the results found in \cite{Barnich:2014kra,Barnich:2015uva}, we show that the Maxwell group in three space-time dimensions can be formulated as a special type of extended semi-direct product based on the $\Sl$ group. This construction can be generalized to define higher-spin extensions of the Maxwell symmetry in 2+1 dimensions, as well as infinite-dimensional enhancements of the $\BMS$ group and the loop group $\LISl$. By including central extensions, we construct Maxwell-like extensions of the  $\hatLISl$ Kac-Moody group and the $\hatBMS$ group, whose associated Lie algebras are exactly the ones found in \cite{Caroca:2017onr}.  The geometric actions associated to these extended semi-direct products are shown to lead to novel  Wess-Zumino terms. They are naturally globally invariant under the considered symmetry groups, and generalize the actions previously obtained in \cite{Alekseev:1988ce,Delius:1990pt,Alekseev:1993qs}(for a recent discussion in the context of three-dimensional gravity see \cite{Rebolledo2015,Barnich:2017jgw}). 

The paper is organized as follows: In Section \ref{sec:esp} we introduce extended semi-direct groups, define the adjoint and coadjoint representations, and construct the geometric action on coadjoint orbits. In Section \ref{sec:maxwell2+1} we show how the Maxwell group fits in this description. In the 2+1-dimensional case, we generalize the results to obtain Maxwell extensions of higher-spin symmetries, loop groups and the $\BMS$ group. In section \ref{centralext}, we define central extensions of extended semi-direct products, and present two novel Wezz-Zumino terms based on the  Maxwell extension of the Kac-Moody and the $\hatBMS$ group. In section \ref{sec:discussion} we conclude with a discussion and possible applications of our results.

\section{Extending semi-direct products}
\label{sec:esp}

In this section we define a particular extension of semi-direct products, which will be used later to formulate the Maxwell group in 2+1 dimensions and its generalizations. Naturally, the starting point in our construction is a semi-direct product group
$K\ltimes_{\sigma}\cv$, where $K$ is a Lie group, $\cv$ is a vector space and $\sigma:K\rightarrow {\rm GL}(\cv)$ is a representation
of $K$ on $\cv$. The elements of a semi-direct product group are denoted
by pairs $\left(U,\alpha\right)$, where $U\in K$ and $\alpha\in \cv$,
and the group operation is given by \cite{rawnsley1975representations,Baguis:1997yp}
\begin{equation}\eqlabel{sdpg}
\left(U,\a \right)\bullet\left(V,\b\right)=\left(U\cdot V \,,\;\a+\sigma_{U}\b\right)\,,
\end{equation}
where $U\cdot V$  denotes the product in $K$. Groups with this structure are ubiquitous in physics. Some examples are
the Euclidean, the Galilei and the Poincar\'e groups \cite{loebl1971group}, as well as the ${\rm BMS}$ group in three and four dimensions\footnote{For a detailed analysis of the rigidity and stability of the $\bmsalg$ algebra in three and four dimensions based on their semi-direct sum structure, see \cite{Parsa:2018kys,Safari:2019zmc}.} \cite{Geroch:1971mz,Barnich:2014kra}.

\subsection{Extended semi-direct product groups}
\label{sec:esdpg}

In order to extend the semi-direct product group \eqref*{sdpg}, we introduce a second vector space $\cvt$, together with a skew-symmetric bilinear map $\times:\cv\times \cv\rightarrow \cvt$, and a representation $\rho:K\rightarrow {\rm GL}(\cvt)$ of $K$ on $\cvt$ compatible with $\sigma$ in the sense that
\be\eqlabel{rho}
\rho_U\left(\a \times \b \right)=\sigma_U \a \times \sigma_U \b \qquad\forall\, \a,\b \in \cv\,.
\ee 
Then, we define the extended group $\cv_{{\rm ext}}=\cv\otimes \cvt$
with elements $\left(\a, a \right)$, $\a \in \cv$, $a \in \cvt$, and the following product:
\begin{equation}\eqlabel{exttranslations}
\left(\a,a\right)\hat{+}\left(\b,b \right)=\left(\a+\b\,,\; a+b +\frac{1}{2}\,\a\times \b \right)\,.
\end{equation}
Note that this operation is non-commutative, and therefore $\cv_{{\rm ext}}$ endowed with the product $\hat{+}$ does not define an Abelian group. Indeed, $\cv_{{\rm ext}}$ defines a central extension of $\cv$, as the function $\cC(\a,\b)=\frac{1}{2}\a\times \b$
is a $\cvt$-valued two-cocycle on $\cv$, i.e.
\begin{equation}
\cC(\a+\b,\gamma)+\cC(\a,\b)=\cC(\a,\b+\gamma)+\cC(\b,\gamma)\,.
\end{equation}
Since $\cv$ and $\cvt$ are Abelian groups, they are respectively isomorphic to their corresponding Lie algebras. Thus, the Lie algebra associated to $\cv_{{\rm ext}}$ is given by the direct sum $\cv\oplus \cvt$ with Lie bracket $\left[\left(\alpha,a\right),\left(\beta,b\right)\right]=\left(0,\,\a\times \b\right)$.

Now we define an \emph{extended semi-direct product} as the group
\begin{equation}
G=K\ltimes_{(\sigma\times\rho)}\cv_{{\rm ext}}\,,\eqlabel{sdp}
\end{equation}
where $\sigma\times\rho$ is the direct product representation
\be
(\sigma\times\rho)_U\left(\a,a\right) = \left(\sigma_U \a\;,\;\rho_U a\right)\,.
\ee
It is important to note that, even though $K$ acts on a central extension
of $\cv$, the group $G$ defines a non-central extension of the original semi-direct product $K\ltimes_{\sigma}\cv$.
The elements of $G$ are given by triplets
$\left(U,\a,a\right)$, where $U\in K$, $\a\in \cv$, and $a \in \cvt$.
The group operation is the natural extension of \eqref*{sdpg},
where now the term $\a+\sigma_{U}\b$ must be replaced by the pair $\left(\a,a\right) \hat{+} \;(\sigma\times\rho)_{U}\left(\b,b\right)$. Using \eqref{exttranslations}, this leads to
\begin{equation}\eqlabel{groupop1}
\left(U,\a,a\right)\bullet\left(V,\b,b\right)=\left(U\cdot V\,,\;\a+\sigma_{U}\b,\; a+\rho_{U} b +\frac{1}{2}\,\a \times\sigma_{U}\b \right).
\end{equation}

\subsubsection*{Extended semi-direct sum algebra}

The Lie algebra associated to $G$ is the \emph{extended semi-direct sum}
\be
\mfg=\mfk\oright_{d\sigma\times d\rho}\cv_{{\rm ext}}\,,
\ee
where $\mfk$ denotes the Lie algebra of $K$, while $d\sigma$ and $d\rho$
stand for the derivative representations of $\sigma$ and $\rho$, respectively.
Denoting the elements of $\mathfrak{g}$
by triplets of the form $\left(X,\a,a\right)$, with $X \in \mfk$, the adjoint action
of $G$ on $\mathfrak{g}$ can be found by evaluating 
\begin{equation}  \eqlabel{Adesp}
{\rm Ad}_{\left(U,\b,b\right)}\left(X,\alpha,a\right)=\left.\frac{d}{d\lambda}\left[\left(U,\b,b\right)\bullet\left(e^{\lambda X},\lambda\a,\lambda a \right)\bullet\left(U,\b,b\right)^{-1}\right]\right|_{\lambda=0} \,.
\end{equation}
The identity of $G$ is $\left(e,0,0\right)$,
where $e$ is the identity in $K$. Thus, the inverse of an element
$\left(U,\a,a \right)$ is given by $\left(U,\a,a \right)^{-1}=\left(U^{-1},-\sigma_{U^{-1}}\a,-\rho_{U^{-1}}a \right)$.
This leads to
\be\eqlabel{AdESP1}
\ba
\Ad_{\left(U,\b, b\right)}\left(X,\alpha,a\right)=\\[6pt]
=\bigg(\Ad_{U}A,\;\sigma_{U}\a-d\sigma_{\Ad_{U}A}\,\b,\;\rho_{U}a-d\rho_{	\Ad_{U}A}\,b +\b \times\sigma_{U}\a-\dfrac{1}{2}\b\times d\sigma_{\Ad_{U}A}\,\b\bigg)\,,
\ea
\ee
where $\Ad _U$ denotes the adjoint representation of $K$. 

The \emph{bracket} on $\mfg$
can be found by looking at the infinitesimal adjoint action, i.e.
\be
\left[\left(X,\alpha,a\right),\left(Y,\beta,b\right)\right]={\rm ad}_{\left(X,\alpha,a\right)}\left(Y,\beta,b\right)=\left.\frac{d}{d\lambda} {\rm Ad}_{\left(e^{\lambda X},\lambda\a,\lambda a\right)}\left(Y,\b,b\right)    \right|_{\lambda=0}\,.
\ee
This defines the adjoint representation of the Lie algebra $\mfg$ in terms of the adjoint representation of $\mfk$, $\ad_X$, and the derivative representations $d\sigma$ and $d\rho$,
\begin{equation}\eqlabel{adESP}
\left[\left(X,\alpha,a\right),\left(Y,\beta,b\right)\right]=\Big(\,\ad_X Y\,,\;d\sigma_{X}\b-d\sigma_{Y}\a\,,\;d\rho_{X}b-d\rho_{Y} a+\a\times \b\Big)\,.
\end{equation}
The elements of the first two slots on the right-hand side of \eqref{AdESP1,adESP} reproduce the form of the adjoint representation of a semi-direct product group and its Lie algebra bracket, respectively (see \cite{Barnich:2014kra,Barnich:2015uva}).
The third slot, however, implies the presence of more Lie algebra generators and
modifies the structure of the usual semi-direct sum algebra.

\subsection{Coadjoint orbits}

Let us consider now the dual space $\mfg^*$. Its elements
will be denoted by triplets of the form $\left(J,\Pi, F\right)$, where $J$ is an element
of the dual space $K^{*}$, while $\Pi$ and $F$ are elements of the dual vector spaces
$\cv^{*}$ and $\cvt^{*}$, respectively. The pairing
between $\mfg$ and its dual space can be naturally constructed
out of the corresponding pairings defined in $\mathfrak{k}$, $\cv$ and $\cvt$
as 
\begin{equation}\eqlabel{pairingesdp}
\left\langle \left(J,\Pi,F\right),\left(X,\a,a\right)\right\rangle_\mfg =\left\langle J,X\right\rangle _{\mfk}+\left\langle \Pi,\a\right\rangle _{\cv}+\left\langle F,a \right\rangle _{\cvt}\,.
\end{equation}
Given a point $\left(J^{0},\Pi^{0},F^{0}\right)\in\mfg^*$, a
coadjoint orbit of the group $G$ can be defined as
\be\eqlabel{orbit1}
O_{\left(J^{0},\Pi^{0},F^{0}\right)}=\left\{ \left(J,\Pi,F\right)={\rm Ad}_{\left(U,\a, a\right)}^{*}\left(J^{0},\Pi^{0},F^{0}\right),\left(U,\a,a\right)\in G\right\} \,,
\ee
where $\Ad_{\left(U,\a, a \right)}^{*}$ denotes de coadjoint action of $G$ on $\mfg^*$. The space $O_{\left(J^{0},\Pi^{0},F^{0}\right)}\subset \mfg^*$ is isomorphic to $G/\,G_{\left(J^{0},\Pi^{0},F^{0}\right)}$, where
$G_{\left(J^{0},\Pi^{0},F^{0}\right)}$ is the
isotropy group of the orbit representative, i.e. 
\be
G_{\left(J^{0},\Pi^{0},F^{0}\right)}= \left\{ \left(U,\a,a\right)\in G\;/\; \Ad_{\left(U,\a, a\right)}^{*}\left(J^{0},\Pi^{0},F^{0}\right)=\left(J^{0},\Pi^{0},F^{0}\right)\right\}\,.
\ee

\subsubsection*{Coadjoint representation}

The coadjoint action of $G$ on $\mfg^{*}$ must be such that the pairing \eqref*{pairingesdp} remains invariant under transformations on the orbit
\be \eqlabel{coAddef}
\left\langle \Ad_{\left(U,\a, a\right)}^{*}\left(J,\Pi,F\right), \Ad_{\left(U,\a, a\right)}\left(X,\a,a\right)\right\rangle_\mfg =\left\langle \left(J,\Pi,F\right),\left(X,\a,a\right)\right\rangle_\mfg \,.
\ee
In order to explicitly write down the coadjoint action of $G$ on its Lie algebra dual, it is necessary to define three extra maps. The first two are the bilinear products $\odot:\cv\times \cv^*\rightarrow \mfg^*$, and $\circledcirc: \cvt\times \cvt^*\rightarrow \mfg^*$, such that
\begin{equation}\eqlabel{prodcirc}
\left\langle \a\odot \Pi,X\right\rangle _{\cv}=\left\langle \Pi,d\sigma_{X} \a\right\rangle _{\cv}\,,\quad
\left\langle a \circledcirc F,X\right\rangle _{\cvt}=\left\langle F,d\rho_{X} a \right\rangle _{\cvt}\,.
\end{equation}
Note that the definition of $\odot$ is familiar from the well-known semi-direct product construction \cite{rawnsley1975representations,Baguis:1997yp,Barnich:2015uva}, while $\circledcirc$ is simply the analogue definition for $\cvt$. The third map necessary to construct the coadjoint representation of an extended semi-direct-product group $G$ is a product $\vee:\cv \times \cvt^*\rightarrow \cv^*$, dual to the map $\times$ in the sense that
\be\eqlabel{dualwedge}
\left\langle  F, \a\times \b \right\rangle_{\cvt} = -\left\langle  \a \vee F,  \b \right\rangle _{\cv}\,.
\ee
These definitions, together with \eqref{AdESP1}, lead to the following form for the coadjoint representation of an extended semi-direct product group:
\be\eqlabel{coAdesdp}
\ba
\Ad_{\left(U,\a, a \right)}^{*}\left(J,\Pi,F\right)=\\[6pt]
=\bigg(\Ad_{U}^{*}J+ \a \odot \sigma_{U}^{*}\Pi  + a \circledcirc\rho_{U}^{*}F
+ \dfrac{1}{2}\a \odot \left(\a \vee \rho_{U}^{*}F \right)\,,
\;\sigma_{U}^{*}\Pi+ \a \vee\rho_{U}^{*}F\,,\;\rho_{U}^{*}F\bigg)\,,
\ea
\ee
where we have used the identity $ \sigma_{U} d\sigma_X \sigma_{U^{-1}} =d\sigma_{\Ad_{U}X}$. Here $\Ad^*_U J$ stands for the coadjoint action of the group $K$ on its Lie algebra dual $\mfk^*$, while $\sigma^{*}$ and $\rho^{*}$ denote the dual
representations of $K$ on $\cv^{*}$ and $\cvt^{*}$, respectively. The definition of the infinitesimal adjoint action follows from differentiating \eqref{coAddef} and yields
\be\eqlabel{coaddef}
\left\langle \ad_{\left(X,\alpha,a\right)}^{*}\left(J,\Pi,F\right), \left(Y,\beta,b\right)\right\rangle_\mfg
+
\left\langle \left(J,\Pi,F\right), \ad_{\left(X,\alpha,a\right)}\left(Y,\beta,b\right)\right\rangle_\mfg
=
0\,.
\ee
Since the Lie bracket of the extended semi-direct sum algebra $\mfg$ is given by \eqref*{adESP}, \eqref{coaddef} can be solved to give
\be
\ad_{\left(X,\alpha,a\right)}^{*}\left(J,\Pi,F\right)=\Big(\ad_{X}^{*}J+\a\odot \Pi+ a \circledcirc F\;,\,\;d\sigma_{X}^{*}\Pi+\a\vee F \;,\,\;d\rho_{X}^{*}F\Big)\,,
\ee
where $d\sigma^{*}$ and $d\rho^{*}$ are dual derivative representations.

\subsubsection*{Geometric action }
Coadjoint orbits are naturally endowed with a symplectic structure \cite{kirillov1976elements,kostant1970orbits,Souriau1970},
which allows one to define a geometric Hamiltonian action with a well-defined
Poisson structure \cite{Alekseev:1988ce,Delius:1990pt,Alekseev:1993qs,Alekseev:2018ful}. In our notation, the Kirillov-Kostant-Souriau symplectic form on a coadjoint orbit $O_{\left(J^{0},\Pi^{0},F^{0}\right)}$ is given by
\be
\Omega\left(v_{(X,\a,a)} , v_{(Y,\b,b)} \right) =
\Big\langle\left(J^{0},\Pi^{0},F^{0}\right)	 , \left[  \left(X,\alpha,a\right) , \left(Y,\beta,b\right) \right] \Big\rangle_\mfg\,,
\ee
where $v_{(X,\a,a)}$ denotes the vector tangent to the orbit at $(J_0,\Pi_0,F_0)$ 
\be
v_{(X,\a,a)}= \ad^*_{\left(X,\alpha,a\right)} \left(J^{0},\Pi^{0},F^{0}\right)\,.
\ee
It can be shown that the pull-back of $\Omega$ on $G$ by the projection map
\be\eqlabel{proy}
{\rm P}: G \rightarrow O_{\left(J^{0},\Pi^{0},F^{0}\right)}: \left(U,\a, a \right)\mapsto \Ad_{\left(U,\a, a \right)}^{*}\left(J^{0},\Pi^{0},F^{0}\right)
\ee
is given by a locally exact two-form \cite{Alekseev:1993qs}
\begin{equation}\eqlabel{KKsf}
{\rm P}^*\Omega= {\rm d}\mathcal{A}\; , \quad \mathcal{A}=-\left\langle \left(J^{0}, \Pi^{0}, F^{0}\right),\left(\Theta_{\a},\Theta_{\a},\Theta_{a}\right)\right\rangle_\mfg \, ,
\end{equation}
where $ {\rm d}$ is the exterior derivative on the group manifold $G$ and the triplet $\left(\Theta_{U},\Theta_{\a},\Theta_{a}\right)$
corresponds to the left-invariant Maurer-Cartan form on $G$, defined by
\begin{equation}\eqlabel{mceq}
{\rm d}\left(\Theta_{U},\Theta_{\a},\Theta_{a}\right)=-\frac{1}{2}\left[\left(\Theta_{U},\Theta_{\a},\Theta_{a}\right),\left(\Theta_{U},\Theta_{\a},\Theta_{a}\right)\right]\,.
\end{equation}
Here wedge product between differential forms is assumed. The corresponding geometric action then reads \cite{Aratyn:1990dj,Kubo:1994cf}
\be\eqlabel{geomaction}
I_G=-\int_{\Gamma}\mathcal{A} ^*\,,
\ee
where $\Gamma$ is a path in $G$,
\be\eqlabel{path}
\begin{array}{ll}
\Gamma: & \mathcal{I}\subset \mathbb{R} \longrightarrow G\\ 
& \quad\quad t \longmapsto \left(U\left( t\right),\a\left( t\right),a\left( t\right)\right)\,,
\end{array}%
\ee
and $\mathcal{A}^*$ denotes the pull-back of the one form $\mathcal{A}$ to $\Gamma$. As remarked in \cite{Alekseev:1988ce,Barnich:2017jgw}, one can define a Hamiltonian $\mathcal{H}$  as a gauge invariant function on the orbit, and include it in the geometric action by adding the term $ - \int dt\, \mathcal{H}$ in \eqref*{geomaction}. The Hamiltonian can be chosen in such a way that the global symmetries of the action are preserved\footnote{The preservation of the symmetries might imply a redefinition of the
gauge parameters of the global symmetries. In \cite{Barnich:2017jgw}, Hamiltonians for
geometric actions are defined in different cases as particular Noether
charges or combinations of them. In \cite{Brzykcy:2017esn}, it is shown how the different
Hamiltonians can be found by looking at the invariant
functions that label the coadjoint orbits.}.

Using \eqref*{adESP}, the Maurer-Cartan equation \eqref*{mceq} can be written as
\be
\ba
{\rm d}\Theta_{U} & = &- \dfrac{1}{2}{\rm ad}_{\Theta_{U}}\Theta_{U}\,,\\
{\rm d}\Theta_{\a} & = & -d\sigma_{\Theta_{U}}\Theta_{\a}\,,\\
{\rm d}\Theta_{a} & = & - d\rho_{\Theta_{U}}\Theta_{a}-\dfrac{1}{2}\Theta_{\a}\times\Theta_{\a}\,.
\ea
\ee
Note that $\Theta _U$ corresponds to the left-invariant Maurer-Cartan form on $K$. Its pull-back to $\Gamma$ reads
\be\eqlabel{defMC}
\Theta_{U}^* = \Theta_{U} \left( \dot{U}(t) \right) \rm{dt}\,,
\ee
where $\dot{U}(t)$ is the tangent vector to the path $U(t)$ in $K$ (with overdot denoting derivative with respect to $t$). Following \cite{Oblak:2017ect}, the action $\Theta_{U}$  on a vector is given by
\be\eqlabel{defMC2}
\Theta_{U} \left( \dot{U}(t) \right)= \left. \frac{d}{d\lambda}\left(U(t)^{-1} U(\lambda)  \right) \right|_{\lambda=t}\,.
\ee
Extending this definition to the full path in $G$ given by \eqref*{path}, with tangent vector $( \dot U(t), \dot \a(t), \dot a(t) )$, we can use the product law \eqref*{groupop1} to find the general expressions
\be\eqlabel{mcform}
\ba
\Theta_{U} \left( \dot U, \dot \a, \dot a \right) &=& \Theta_{U} \left( \dot{U}(t) \right)\,,\\[5pt]
\Theta_{\a}  \left( \dot U, \dot \a, \dot a \right) &=&   \sigma_{U(t)^{-1}} \dot \a (t)\,, \\[5pt]
\Theta_{a}  \left( \dot U, \dot \a, \dot a \right) &=& \rho_{U(t)^{-1}}\left( \dot a (t) - \dfrac{1}{2} \a (t)\times \dot \a (t)  \right)\,.
\ea
\ee
Using \eqref*{mcform} in \eqref{KKsf}, the action \eqref*{geomaction} can be written as
\be\eqlabel{espga}
I_G= \int dt \left[ \left\langle J^{0}, \Theta_{U}(\dot U) \right\rangle_{\mfk} + \left\langle \sigma^* _U \Pi^{0}, \dot\a \right\rangle_{\cv} + \left\langle \rho^* _U F^{0}, \left( \dot a-\frac{1}{2}\a \times \dot \a \right) \right\rangle_{\cvt} \right]\,.
\ee
The geometric action \eqref*{espga} is naturally globally invariant under left transformations of the form $(U,\a,a)\rightarrow (V\cdot U,\b+\sigma_V \a, b+\sigma_V a +\frac{1}{2} \b\times\sigma_V\a )$, while gauge symmetries are given by right transformations $(U,\a,a)\rightarrow (U\cdot V(t),\a+\sigma_U \b(t), a+\sigma_U b(t) +\frac{1}{2} \a\times\sigma_U\b(t) )$, provided $(V(t),\b(t),b(t))\in G_{(J^0,\Pi^0,F^0)}$, as shown in \cite{Barnich:2017jgw}\footnote{Note that the form of the geometric action \eqref*{geomaction} is equivalent to the general definition given in \cite{Barnich:2017jgw} when setting $g^{-1}=(U,\a,a)$. When written in terms of $g$, the role of right and left transformations are interchanged.}.

\subsection{Extended semi-direct products under the adjoint action}
\label{extsdpAd}

Several semi-direct product groups that define symmetries of $2+1$ dimensional systems have the particular structure $K\ltimes_{\Ad} \mfk^{\rm (ab)}$, where $\mfk^{\rm (ab)}$ denotes the Lie algebra of $K$ seen as a vector space, and $K$ acts on its Lie algebra by the adjoint action. Examples are the $ISO(2,1)$ group, the $\hatBMS$ group, and higher-spin extensions thereof \cite{Barnich:2014kra,Campoleoni:2016vsh} (for a detailed analysis, see also \cite{Oblak:2016eij}).  Here we show how to generalize this special kind of symmetries to the case of extended semi-direct product groups, which will be the key construction to define in a unified manner the different Maxwell-like symmetries in 2+1 dimensions that will be treated in the subsequent sections.

Let us consider a particular case of extended semi-direct product \eqref*{groupop1}, where $\cv$ and $\cvt$ are given by the Lie algebra associated to the Lie group $K$ seen as an Abelian vector group, and the representations $\sigma$ and $\rho$ are given by the adjoint representation of $K$, i.e.
\be\eqlabel{special}
\cv=\cvt=\mfk^{\rm (ab)}\, ,\quad \sigma=\rho=\Ad\,.
\ee
In this case, the requirement \eqref*{rho} is trivially satisfied provided the anti-symmetric bilinear map $\times$ is chosen as
\be\eqlabel{xprodad}
\a \times \b=\mbox{{\rm ad}}_{\a}\b\,,\quad\forall \a,\b\in \mfk\,.
\ee
With these considerations, we can define the extended vector space $\mfk^{\rm (ab)}_{\rm ext}$, given by the set $\mfk^{\rm (ab)}\times\mfk^{\rm (ab)}$, and endowed with a group law of the form \eqref*{exttranslations}. Thus, we can define the \emph{special} extended semi-direct product 
\be\eqlabel{esdpad}
H=K\ltimes_{{\rm Ad}}\mfk^{\rm (ab)}_{\rm ext} \,,
\ee
where the group operation follows from \eqref{groupop1}, and is given by 
\begin{equation}\eqlabel{groupop2}
\left(U,\a, a \right)\bullet\left(V,\b,b\right)=\left(U\cdot V,\; \a+\Ad_{U}\b\,,\; a +\Ad_{U} b +\frac{1}{2}\ad_\a \Ad_{U}\b   \right)\,.
\end{equation}
As in this case the derivative representations $d\sigma$ and $d\rho$ are both given by the adjoint representation of $\mfk$, the general form for the adjoint representation of an extended semi-direct product \eqref*{AdESP1} can be easily evaluated to give
\be\eqlabel{AdESP2}
\ba
\Ad_{\left(U,\b, b \right)}\left(X,\alpha,a\right)=\\[6pt]
=\bigg({\rm Ad}_{U}X,\,{\rm Ad}_{U}\a-\ad_{\Ad_{U}X} \b,\,\Ad_{U} a-\ad_{\Ad_{U}X} b
+\ad_\b \left(\Ad_{U} \a-\frac{1}{2}\ad_{\Ad_{U}X}\b   \right)    \bigg)\,.
\ea
\ee
Differentiating this expression leads to the Lie algebra bracket associated to the corresponding Lie algebra
\be\eqlabel{sesdpa}
\mfh=\mfk \oright _\ad \mfk_{\rm ext}^{({\rm ab})} \,,
\ee
which can also be directly obtained from the general expression  \eqref*{adESP}:
\be\eqlabel{adESP2}
\left[\left(X,\alpha,a\right),\left(Y,\beta,b\right)\right]=\Big( \ad_X Y  ,\;\ad_X \b -\ad_Y \a ,\;\ad_X b -\ad_Y a + \ad_\a \b \Big)\,.
\ee

Coadjoint orbits are defined according to \eqref{orbit1}, where the form of the coadjoint representation follows from \eqref*{coAddef}. In the case of a special extended semi-direct product, the pairing \eqref*{pairingesdp} for the algebra $\mfh$ is written solely in terms of the paring in $\mfk$,
\be\eqlabel{pairingesdp2}
\left\langle \left(J,\Pi,F\right),\left(X,\alpha,a\right)\right\rangle_\mfh =\left\langle J,X\right\rangle _{\mfk}+\left\langle \Pi,\a\right\rangle _{\mfk}+\left\langle F,a\right\rangle _{\mfk}\,,
\ee
while \eqref{special} implies that $\sigma^*=\rho^*=\ad^*$. Therefore \eqref*{prodcirc,dualwedge} lead to
\be
\a\odot \pi = \a \circledcirc \b = \a \vee \b =\ad^*_\a \b \,.
\ee
Putting this together allows one to write down the coadjoint representation of $H$ directly from \eqref*{coAdesdp}, yielding
\be\eqlabel{adcoad2}
\ba
\Ad_{\left(U,\a, a \right)}^{*}\left(J,\Pi,F\right)=\\[6pt]
=\bigg(\Ad_U^* J+\ad_\a^*\left(\Ad_U^*\Pi+\frac{1}{2}\ad_\a^*\Ad_U^* F \right) +\ad_a^*\Ad_U^*F,\,\Ad_{U}^{*}\Pi+\ad_{\a}^{*}\Ad_{U}^{*}F ,\,\Ad_{U}^{*}F\bigg)\,,
\ea
\ee
while its infinitesimal form is given by
\be\eqlabel{infcoad}
\ad_{\left(X,a,\text{\ensuremath{\alpha}}\right)}^{*}\left(J,\Pi,F\right)=\Big(\ad_{X}^{*}J+{\rm ad}_{\a}^{*}\Pi+\ad_{a}^{*}F\,,\;\ad_{X}^{*}\Pi+\ad_{\a}^{*}F\,,\;\ad_{X}^{*}F\Big) \,.
\ee
The construction of the geometric action follows from \eqref{KKsf,geomaction} by means of \eqref*{special,xprodad}. This gives
\be\eqlabel{espga2}
I_H= \int dt\left[ \left\langle J^{0}, \Theta_{U}(\dot U) \right\rangle_{\mfk} + \left\langle \Ad^* _U \Pi^{0}, \dot\a \right\rangle_{\mfk} + \left\langle \Ad^* _U F^{0}, \left(\dot a-\frac{1}{2}\ad_\a  \dot\a\right) \right\rangle_{\mfk} \right]\,.
\ee

\section{Maxwell groups}
\label{sec:maxwell2+1}

\subsection{The Maxwell group in $D$ dimensions}

The Maxwell group in $D$ dimensions, which will be denoted by $\Maxd$, can be constructed by generalizing the original definition for $D=4$, given in \cite{Schrader:1972zd}. It can be written as an extended semi-direct product \eqref*{sdp}, where $K=SO(D-1,1)\uparrow$ is the ortochronus Lorentz group in $D$ dimensions, $\cv=\mathbb{R}^D$ is the vector space of translations, and $\cvt=\so\, (D) \cong \bigwedge^2 \left(\mathbb{R}^D\right)$ is the space of $D\times D$ anti-symmetric matrices, isomorphic to the second exterior power of $\mathbb{R}^D$. Therefore, the Maxwell group is given by
\be\eqlabel{maxwellgroup}
\Maxd =SO(D-1,1)\uparrow \ltimes \mathbb{R}_{\rm ext}^{D} \,,
\ee
where  $\mathbb{R}_{\rm ext}^{D}$ is a central extension of the translation group $\mathbb{R}^D$ by $\bigwedge^2 \left(\mathbb{R}^D\right)$.  As seen in Section \ref{sec:esdpg}, this extension requires a skew-symmetric bilinear product $\times: \mathbb{R}^D\times \mathbb{R}^D \rightarrow  \bigwedge^2 \left(\mathbb{R}^D\right)$. By choosing an orthonormal basis $e_\mu$ ($\mu=1,\ldots,D$) on $\mathbb{R}^D$, the product $\times$ can be constructed by means of the usual \emph{cross product}
\be\eqlabel{crossmaxwell}
e_\mu \times e_\nu =  e_\mu  \otimes e_\nu -  e_\nu  \otimes e_\mu \,,
\ee
which defines the natural basis for $\bigwedge^2 \left(\mathbb{R}^D\right)$. Then, the extended group of translations $\mathbb{R}_{\rm ext}^{D}$ is the centrally extended group with elements $\left(\a, a \right)$, $\a\in\mathbb{R}^D$, $a \in\bigwedge^2 \left(\mathbb{R}^D\right)$, endowed with a product $\hat{+}$ of the form \eqref*{exttranslations} with
\be\eqlabel{cprodmax}
\a=\a^\mu e_\mu \;, \quad a=\frac{1}{2}a^{\mu\nu}e_\mu\times e_\nu \;, \quad \a\times \b =\frac{1}{2}\left( \a^\mu \b^\nu  -  \a^\nu \b^\mu \right) e_\mu \times e_\nu \,.
\ee
The action of a Lorentz transformation $\Lambda\in SO(D-1,1)\uparrow$ on the elements of $\mathbb{R}^D$ and $\bigwedge^2 \left(\mathbb{R}^D\right)$ are given by the standard vector and tensor representations, respectively
\be\eqlabel{repLorentz}
\sigma_\Lambda \a \equiv \Lambda \a = \Lambda^{\mu} _{\;\;\nu} \a^\nu e_\mu \;,\quad \rho_\Lambda a \equiv \Lambda a \Lambda^T =\frac{1}{2} \Lambda^\mu _{\;\;\alpha} \Lambda^\nu _{\;\;\beta} a^{\alpha\beta} e_\mu \times e_\nu \,.
\ee
Note that, in this case, $\rho$ corresponds to the exterior power representation $\rho= \sigma_{\bigwedge^2\left({\mathbb{R}^D}\right)}$. Therefore, it automatically satisfies \eqref{rho}.  Putting all this together, the product law of the ${\rm Maxwell}_D$ group follows directly from \eqref{groupop1}. Denoting its elements by triads of form $\left(\Lambda, \a, a \right)$, we get
\begin{equation}\eqlabel{groupopmaxwell}
\left(\Lambda, \a, a \right)\bullet\left(\Sigma,\b, b\right)=\left(\Lambda \Sigma ,\;\a+\Lambda \b,\;  a+\Lambda b \Lambda^T+\frac{1}{2}\,\a\times\Lambda \b \right).
\end{equation}

The Maxwell algebra is given by the extended semi-direct sum 
\be\eqlabel{maxalgd}
\maxd =\so\,(D-1,1)\oright \mathbb{R}_{\rm ext}^{D}\,.
\ee
As both $\mathbb{R}^D$ and $\bigwedge^2 \left(\mathbb{R}^D\right)$ are Abelian vector groups, their corresponding Lie algebras are isomorphic to themselves. This means we can define the generators
\be\eqlabel{genPiF}
\bPi_\mu = \left(0,e_\mu,0\right) \;,\quad \bF_{\mu\nu}= \left(0,0,e_\mu \times e_\nu\right) \, ,
\ee
so that the elements of the $\maxd$ algebra can be defined from the following exponential map
\be
\left(X, \a , a \right)=\frac{1}{2} X^{\mu\nu} \bJ_{\mu\nu}+ \a^\mu \bPi_\mu + \frac{1}{2}a^{\mu\nu} \bF_{\mu\nu} = \frac{d}{d\lambda} \left. \left( e^{\lambda X} , \lambda \, \a^\mu e_\mu \, , \frac{\lambda}{2} a^{\mu\nu} e_\mu \times e_\nu \right) \right|_{\lambda=0} \,,
\ee
where $\bJ_{\mu\nu}$ stands for the generators of Lorentz transformations
\be\eqlabel{genJ}
\bJ_{\mu\nu}\equiv\left(\cJ_{\mu\nu},0,0\right)\;,\quad 
\ee
and $\cJ_{\mu\nu}$ stands for the usual matrix representation for $\so\,(D-1,1)$:
\be
 \left( \cJ_{\mu\nu} \right)^\rho _{\;\;\sigma}=\delta_\mu^\rho\eta_{\nu\sigma}-\delta_\nu^\rho \eta_{\mu\sigma}\,.
\ee
The derivative representations $d\sigma$ and $d\rho$ are given by the infinitesimal limit of \eqref{repLorentz}, 
\be\eqlabel{derrepLorentz}
d\sigma_X \a = \frac{1}{2}X^{\mu\nu} \left( \cJ_{\mu\nu} \right)^\rho _{\;\;\sigma} \a^\sigma e_\rho \;,\quad
d\rho_X a =  \frac{1}{4}  X^{\mu\nu}    \left[
 \left( \cJ_{\mu\nu} \right)^\rho _{\;\;\lambda} a ^{\lambda\sigma}
+
 \left( \cJ_{\mu\nu} \right)^\sigma _{\;\;\lambda} a^{\rho\lambda}
\right]    e_\rho \times e_\sigma \,.
\ee 
The definitions \eqref*{cprodmax,repLorentz,derrepLorentz}, plus the fact that the adjoint action of the Lorentz group on its Lie algebra is given by $\Ad_\Lambda X= \Lambda X \Lambda^T$, enables one to write down the adjoint representation of the Maxwell group using expression \eqref*{AdESP1}. Similarly, the bracket for the $\maxd$ algebra follows from \eqref*{adESP}.  When applied to the generators \eqref*{genPiF,genJ}, it leads directly to the commutation relations \eqref*{maxwellalg}.

\subsubsection*{Particle action}

The $\maxd$ algebra and its dual space are isomorphic. The isomorphism can be explicitly constructed by endowing $\mathbb{R}^D$ with the  $D$-dimensional Minkowski metric $\eta_{\mu\nu}={\rm diag}\left(-1,1,\dots,1\right)$, which maps contravariant vectors and tensors into covariant ones. The pairing \eqref*{pairingesdp} can be constructed considering
\be
\ba
&&\left\langle J,X\right\rangle _{\mfk} = \dfrac{1}{2}J_{\mu\nu} X^{\mu\nu} = \dfrac{1}{2}\eta_{\mu_\rho}\eta_{\nu\sigma} J^{\mu\nu}X^{\rho\sigma} \,,  \\ [6pt]
&&\left\langle \Pi,\a\right\rangle _{\cv} = \Pi_\mu \a^\mu= \eta_{\mu\nu}\Pi^\mu \a^\nu  \,, \\ [6pt]
&&\left\langle F,a \right\rangle _{\cvt} =  \dfrac{1}{2}F_{\mu\nu} a^{\mu\nu} = \dfrac{1}{2}\eta_{\mu_\rho}\eta_{\nu\sigma} F^{\mu\nu} a^{\rho\sigma} \,.
\ea
\ee
The coadjoint representation of the Maxwell group and the Maxwell algebra can be obtained from \eqref*{adcoad2} and \eqref*{infcoad}, respectively, where the dual representations $\sigma^*$ and $\rho^*$ as well as their corresponding derivative representations are given by the covariant versions of \eqref{repLorentz,derrepLorentz}. The products $\odot$ and $\circledcirc$ in this case can be worked out using \eqref*{prodcirc}, and lead to elements in the Lie algebra dual $\so\,(D-1,1)^*$ with components
\begin{equation}\eqlabel{prodsmaxwell}
\begin{array}{lcl}
\left( \a\odot \Pi\right)_{\mu\nu}&=& \a^\rho \left(\eta_{\nu\rho}\Pi_{\mu}-\eta_{\mu\rho} \Pi_{\nu}\right)\,,\\[5pt]
\left( a \circledcirc F\right)_{\mu\nu}&=& \dfrac{1}{2} a^{\rho\sigma}\left(\eta_{\mu\sigma}F_{\nu\rho}+\eta_{\nu\rho}F_{\mu\sigma}-\eta_{\mu\rho}F_{\nu\sigma}-\eta_{\nu\sigma}F_{\mu\rho}\right)\, ,
\end{array}
\end{equation}
while \eqref*{dualwedge} takes the form
\be\eqlabel{dualwedgemaxwell}
\left(\a\vee F \right)_\mu = F_{\mu\nu}\a^\nu \,.
\ee
Once the coadjoint representation is known, coadjoint orbits can be defined according to \eqref{orbit1}. Then, one can construct a geometric action that is naturally invariant under the Maxwell group. The key ingredient in this construction is the left-invariant Maurer-Cartan form \eqref*{mcform}, which in this case reads
\be\eqlabel{maxwellmcform}
\ba
 \Theta_\Lambda &=& \Lambda^{-1}{\rm d}\Lambda \,, \\[5pt]
\Theta _\a &=& \Lambda^\mu_{\;\;\nu} {\rm d} \a^\nu \bPi_\mu \,,   \\[5pt]
\Theta_ a &=& \dfrac{1}{2}\Lambda^\mu_{\;\;\rho}\Lambda^{\nu}_{\;\;\sigma}\left[ {\rm d} a^{\rho\sigma} -\dfrac{1}{2}\left( \a^{\rho}{\rm d}\a^{\sigma}-\a^{\sigma} {\rm d} \a^{\rho} \right)\right]\bF_{\mu\nu} \,.
\ea
\ee
The Maxwell group satisfies Mackey's theorem \cite{Schrader:1972zd,negro1990local1}, and therefore its irreducible representations can be obtained from the irreducible representations of $\cv_{\rm ext}$. As these representations correspond to Poincar\'e  particles in the presence of a constant electromagnetic field, the Hamiltonian geometric action \eqref*{geomaction} for the Maxwell group is expected to describe world lines of such particles. This is indeed the case. In fact, evaluating \eqref*{espga} for a representative of the form $\left(0,\Pi^0,F^0 \right)\in\cvt^*$, we get
\be\eqlabel{maxwellaction1}
I_{\Maxd}= \int_\Gamma dt \left[(\Lambda \Pi^{0})_\mu\dot \a^\mu  +\frac{1}{2}(\Lambda F^{0} \Lambda^T)_{\mu\nu} \left( \dot a^{\mu\nu}-\frac{1}{2} (\a^\mu  \dot \a ^\nu  -  \a^\nu  \dot \a ^\mu )  \right)  \right]\,.
\ee
The orbit in $\cvt^*$ with representative $\left(\Pi^0,F^0\right)$ is defined as $O_{\left(\Pi^0,F^0\right)}: \left(\pi,f\right)=\left(\sigma_\Lambda \Pi^0, \rho_\Lambda F^0\right)$, therefore we define the \emph{momentum} and the \emph{electromagnetic field} on the orbit as
\be
\pi_\mu = \Lambda^\mu_{\;\;\nu} \Pi^0_\mu \;,\quad f_{\mu\nu}=  \Lambda^\mu_{\;\;\rho}  \Lambda^\mu_{\;\;\sigma} F^0_{\rho\sigma} \,.
\ee
The path $\Gamma\subset \Maxd$ defines a path on the orbit $O_{(0,\Pi_0,F_0)}$ through the projection map \eqref*{proy}. We can consider an orbit of the Maxwell group describing a massive particle of mass $M$. The mass shell condition $\pi_\mu \pi^\nu=-M^2$ can be implemented in \eqref*{maxwellaction1} as a Hamiltonian by means of a Lagrange multiplier $e$. Finally, by identifying the translation components $\alpha^\mu$ with space-time coordinates $x^\mu$ and the elements $a^{\mu\nu}$ with extra coordinates $y^{\mu\nu}=-y^{\nu\mu}$, the geometric action takes the form
\be\eqlabel{maxwellaction2}
I_{\Maxd}= \int dt \left[ \pi_\mu\dot x^\mu  +\frac{1}{2}f_{\mu\nu} \left( \dot y^{\mu\nu}-\frac{1}{2} (x^\mu  \dot x ^\nu  -  x^\nu  \dot x ^\mu )  \right) - \frac{e}{2} \left(\pi_\mu \pi^\mu +M^2 \right) \right]\;.
\ee
This is the action of a massive particle coupled to a constant electromagnetic field in Hamiltonian form, and it has been previously studied from the point of view of non-linear realizations \cite{Bonanos:2008ez,Gibbons:2009me,Bonanos:2010fw}. Here, we have shown that it can obtained from the Kirillov-Kostant-Souriau symplectic structure on coadjoint orbits of the Maxwell group. The elements of the orbit $O_{(0,\Pi_0,F_0)}$ can be computed using \eqref{coAdesdp} together with \eqref*{prodsmaxwell,dualwedgemaxwell}. Their components read
\be \eqlabel{noethermaxwell}
\ba
J_{\mu\nu}&=&x_\mu p_\nu  -x_\nu p_\mu + y_{\mu\rho} f_\nu^{\;\;\;\rho}-y_{\nu\rho} f_\mu^{\;\;\;\rho}\,,
\\[5pt]
\Pi_\mu &=& \pi_\mu + f_{\mu\nu}x^\nu =  p_\mu + \dfrac{1}{2}f_{\mu\nu}x^\nu \,, \\[5pt]
F_{\mu\nu}&=&f_{\mu\nu}\,,
\ea
\ee
where we have defined $p_\mu=\delta I_{Maxwell}/\delta \dot x=\pi_\mu+\frac{1}{2}f_{\mu\nu}x^\nu$ as the canonical momenta associated to $x^\mu$. We see that $f_{\mu\nu}$ corresponds to the momentum associated to the extra coordinates $y^{\mu\nu}$, which play the role of Lagrange multipliers for the constraints $\dot f_{\mu\nu} =0 $. Therefore, $\pi_\mu$ is the momenta that accounts for the minimal coupling to a constant electromagnetic potential 
\be
\pi_\mu= p_\mu+A_\mu  \;,\quad A_\mu= - \frac{1}{2}f_{\mu\nu}x^\mu \,.
\ee
The quantities $J_{\mu\nu}$, $\Pi_\mu$, and $F_{\mu\nu}$ define a basis for the Noether charges \eqref*{noethermaxwell}. By definition, they satisfy a Poisson algebra isomorphic to glogal symmetry action of the algebra \cite{Barnich:2017jgw}, i.e. the $\maxd$ algebra \eqref*{maxwellalg},
\be\eqlabel{maxwellalgPoisson}
\begin{array}{lcl}
\left\{ J_{\mu\nu}, J_{\rho\sigma}\right\}=\eta_{\mu\rho} J_{\nu\sigma}+\eta_{\nu\sigma} J_{\mu\rho}-\eta_{\mu\sigma} J_{\nu\rho}-\eta_{\nu\rho} J_{\mu\sigma}\,, & \quad & \left\{ \Pi_{\text{\ensuremath{\mu}}}, \Pi_{\nu}\right\}=  F_{\mu\nu}\,,\\
\left\{ J_{\mu\nu}, \Pi_{\rho}\right\}=\eta_{\mu\rho} \Pi_{\nu}-\eta_{\nu\rho} \Pi_{\mu}\,, & \quad & \left\{ \Pi_{\mu}, F_{\nu\rho}\right\}=0\,,\\
\left\{ J_{\mu\nu}, F_{\rho\sigma}\right\}=\eta_{\mu\rho} F_{\nu\sigma}+\eta_{\nu\sigma} F_{\mu\rho}-\eta_{\mu\sigma} F_{\nu\rho}-\eta_{\nu\rho} F_{\mu\sigma}\,, & \quad & \left\{ F_{\mu\nu}, F_{\rho\text{\ensuremath{\sigma}}}\right\}=0\,.
\end{array}
\ee
As $p_\mu$ and $f_{\mu\nu} $ are the canonical momenta associated to the coordinates $x^\mu$ and $y^{\mu\nu}$, respectively, one can define a coordinate representation for them as
\be
p_\mu\rightarrow -i\dfrac{\d}{\d x^\mu}\;,\quad f_{\mu\nu}\rightarrow -i\dfrac{\d}{\d y^{\mu\nu}} \,,
\ee
which leads to 
\be\eqlabel{diffmaxwell}
\ba
J_{\mu\nu}&\rightarrow&-i\left(x_\mu \dfrac{\d}{\d x^\nu}  -x_\nu \dfrac{\d}{\d x^\mu}+ y_{\mu}^{\;\;\rho}\dfrac{\d}{\d y^{\nu\rho}}-y_{\nu}^
{\;\;\rho} \dfrac{\d}{\d y^{\mu\rho}}\right) \,,
\\[5pt]
\Pi_\mu &\rightarrow&   -i\left( \dfrac{\d}{\d x^\mu} + \dfrac{1}{2}x^\nu \dfrac{\d}{\d y^{\mu\nu}} \right) \,,  \\[5pt]
F_{\mu\nu}&\rightarrow& -i \dfrac{\d}{\d y^{\mu\nu}} \,,
\ea
\ee
and fulfill \eqref{maxwellalg} when Poisson brackets are replaced by commutators in the form $\{\ \;,\;\}\ \rightarrow i\left[\;,\;\right]$. As stated in \cite{Gibbons:2009me}, the differential operators \eqref*{diffmaxwell} define the Killing vectors fields of a $\left(D+D(D-1)/2\right)$-dimensional superspace with metric
\be\eqlabel{metricmaxwell}
ds^2= \left(\eta_{\mu\nu} +\frac{1}{4}\left(x^\rho x_\rho \eta_{\mu\nu}-x_\mu x_\nu \right)\right)dx^\mu dx^\nu +   \eta_{\mu\nu}x_\rho dx^\mu dy^{\nu\rho}   +\frac{1}{2}\eta_{\mu\rho}\eta_{\nu\sigma}dy^{\mu\nu}dy^{\rho\sigma}  \,,
\ee
and whose isometries are given by
\be
\ba
x^\mu&\rightarrow& \Lambda^\mu_{\;\;\nu}x^\nu+\a^\mu \,,
\\[5pt]
y^{\mu\nu}&\rightarrow&  \Lambda^\mu _{\;\;\rho} \Lambda^\nu _{\;\;\sigma} y^{\rho\sigma}+\dfrac{1}{2}\left(\a^\mu \Lambda^\nu_{\;\;\rho}x^\rho-\a^\nu  \Lambda^\mu_{\;\;\rho}x^\rho \right)+a^{\mu\nu} \,.
\ea
\ee
Note that the Hamiltonian used in \eqref{maxwellaction2} is not the most general one. In fact, it is possible to add more constraints to the action constructed out of the Casimir invariants of the system, which can fix components of the representative element $F^0$, associated to the constant electromagnetic field \cite{Gibbons:2009me}. 

\subsection{The Maxwell group in 2+1 dimensions}

The Maxwell group in 2+1 dimensions is obtained from \eqref{maxwellgroup} for $D=3$, i.e.
\be\eqlabel{maxwellg3}
\Max= {\rm SO}\left(2,1\right) \ltimes \mathbb{R}_{\rm ext}^{3}\,.
\ee
However, we will show now that, unlike its higher-dimensional versions, the $\Max$ group can be alternatively described by the special kind of extended semi-direct product defined in Section \ref{extsdpAd}. This construction will allow us to define different generalizations of the Maxwell symmetry in 2+1 dimensions.

In order to formulate the $\Max$ group as a special extended semi-direct product of the form \eqref*{esdpad}, we will follow the same strategy used in \cite{Barnich:2014kra} for the $ISO(2,1)$ group. First, we will consider the double covering of \eqref*{maxwellg3}:
\be\eqlabel{idv}
\dMax= \Sl \ltimes \mathbb{R}_{\rm ext}^{3} \,.
\ee
Secondly, vectors in $\mathbb{R}^3$ can be identified with elements of the $\sl$ algebra, seen as an Abelian vector group
\be
\alpha=\alpha^\mu e_\mu \longrightarrow \alpha= \alpha^\mu \bt_\mu \,,
\ee
where $\bt_\mu$ denotes the generators of $\sl$,
\be
\bt_1= \frac{1}{2}\begin{pmatrix} 0 & 1  \\ -1 & 0 \end{pmatrix}\,,\quad
\bt_1=  \frac{1}{2}\begin{pmatrix} 0 & 1  \\ 1 & 0 \end{pmatrix}\,,\quad
\bt_1=  \frac{1}{2}\begin{pmatrix} 1 & 0  \\ 0 & -1 \end{pmatrix}\,,
\ee
which satisfy the commutation relations
\be\eqlabel{sl2alg}
\left[\bt_{\mu},\bt_{\nu}\right]=\epsilon^\rho_{\;\;\mu\nu}\bt_{\rho}\,.
\ee
The same can be done with the space $\bigwedge^2 (\mathbb{R}^3)$. In fact, any $\mathfrak{so}(3)$ matrix can be identified with a vector in $\mathbb{R}^3$. In other words, the spaces $\bigwedge^2 (\mathbb{R}^3)$ and $\mathbb{R}^3$ can be shown to be isomorphic by means of Hodge duality. Thus, $\bigwedge^2 (\mathbb{R}^3)$ can also be identified with $\slab$,
\be\eqlabel{idvt}
a=\frac{1}{2}a^{\mu\nu} e_\mu \times e_\nu\longrightarrow a= a^\rho \bt_\rho   \,,\quad a^\rho =\frac{1}{2}\epsilon^{\rho}_{\;\;\mu\nu}a^{\mu\nu}\,.
\ee
With this identification, the inner product of vectors in $\mathbb{R}^3$ is reproduced by the Killing form on $\sl$
\be\eqlabel{killingsl2}
2{\rm Tr}\left[\alpha \beta \right]= \eta_{\mu\nu}\a^\mu \b^\nu\,,
\ee
while the cross product \eqref*{crossmaxwell} is translated into matrix commutation, $\alpha \times \beta =[\alpha,\beta]$. This means that we can make the identification 
\be\eqlabel{idvext}
\mathbb{R}^3_{\rm ext}\rightarrow \slabe=\slab \times \slab\,,
\ee
where $ \slabe$ is endowed with a product of the form \eqref*{exttranslations} given by
\be\eqlabel{prodslext}
\left(\a,a\right)\hat{+}\left(\b,b \right)=\left(\a+\b\,,\; a+b +\frac{1}{2}\, \left[\a, \b\right] \right)\,.
\ee
Lorentz transformations on vectors \eqref*{repLorentz} are realized as the adjoint action of elements $U\in \Sl$ on the $\sl$ algebra \cite{Barnich:2014kra}
\be\eqlabel{ltransformations }
\Lambda \alpha \longrightarrow U \alpha U^{-1}\;,\quad \Lambda^{\mu}_{\;\;\nu}=2\eta^{\mu\rho} {\rm Tr}\left[\bt_\rho U \bt_\nu U^{-1}\right]\,.
\ee
Therefore, the elements of the Maxwell group can be considered as  triplets {$\left(U,\a, a \right)$},
where $U$ is an $SL(2,\mathbb{R})$ matrix and $a$,$\alpha$ are
$\sl$ matrices. Hence, the group operation \eqref*{groupopmaxwell} turns into
\begin{equation}\eqlabel{prodmaxwel2+1}
\left(U,\a, a \right)\bullet\left(V,\b,b\right)=\left(UV ,\; \a+ U \b U^{-1}\,,\; a+  U\, b U^{-1}+\frac{1}{2}\left[\a, U\b U^{-1}\right]\right) \,,
\end{equation}
which matches the structure of \eqref*{groupop2}. This means that the double covering of the Maxwell group in 2+1 dimensions corresponds to a special extended semi-direct product \eqref*{esdpad}, where $K=\Sl$
\begin{equation}\eqlabel{dmaxwell2+1}
\dMax=\Sl\ltimes_\Ad \slabe \,.
\end{equation}
When one copy of the $\slab$
algebras in \eqref*{prodmaxwel2+1} is eliminated, \eqref*{dmaxwell2+1} reduces
to the double covering of the Poincar\'e group in three dimensions, $\Sl\ltimes_\Ad \slab$, which has been studied in \cite{Barnich:2014kra,Barnich:2015uva}.

Following the construction developed in Section \ref{extsdpAd}, the Maxwell algebra in 2+1 dimensions has the form \eqref*{sesdpa} for $\mfk=\sl$, i.e.
\be\eqlabel{maxalgstructure}
\max=\sl\oright_\ad \slabe \,.
\ee
This definition is isomorphic to \eqref*{maxalgd} for $D=3$ due to the isomorphism $\so(2,1)\simeq\sl$ and the identification \eqref*{idvext}. Its bracket follows from \eqref{adESP2} and the commutation relations of the $\sl$ algebra given in \eqref*{sl2alg}. As $\Sl$ is a matrix group, the Lie algebra $\sl$ and its dual space can be identified by means of the Killing form \eqref*{killingsl2}.  A base for $\sl^*$ is given by $\bt^\mu=\eta^{\mu\nu}\bt_\nu$ and therefore any element in $\sl$ can be written as an element of $\sl^*$ and vice-versa. Thus, the pairing between $\sl$ and its dual space is naturally given by
\be
\left\langle J, X \right\rangle_{\sl} =  J_\mu X^\nu \tr \left[\bt_\mu \bt^\nu\right] = \frac{1}{2}J_\mu X^\mu = \frac{1}{2} \left\langle  J, X  \right\rangle_{\mathbb{R}^3} \, .
\ee
The adjoint and the coadjoint action are equivalent and given by matrix conjugation, and their infinitesimal limits correspond to matrix commutation,
\be\eqlabel{adcoadsl}
\Ad_{U}X=U X U^{-1}\, ,\quad
\Ad_{U}^{*}J=U J U^{-1}\, , \quad
\ad_{X}Y=\epsilon^\rho_{\;\;\mu\nu}X^\mu Y^\nu\bt_\rho \; , \quad
\ad_{X}^{*}J=\epsilon^\nu_{\;\;\rho\mu}X^\mu J_\nu\bt^\rho \, .
\ee
This allows one to construct the adjoint and the coadjoint representations of the Maxwell group in 2+1 dimensions and its Lie algebra from expressions \eqref*{AdESP2,adcoad2}, respectively, together with their infinitesimal forms \eqref*{adESP2,infcoad}.

\subsection{The Maxwell algebra in 2+1 dimensions}
\label{maxalgebra2+1}

The explicit form of the Maxwell algebra in 2+1 dimensions can be written down by defining the following generators
\be\eqlabel{genmax3d}
\bJ_{\mu} \equiv\left(\bt_{\mu},0,0\right)\,, \quad
\bPi_{\mu} \equiv\left(0,\bt_{\mu},0\right)\,, \quad
\bF_{\mu} \equiv\left(0,0,\bt_{\mu}\right)\,.
\ee
where $\bt_\mu$ are the generators of the $\sl$ algebra and satisfy \eqref*{sl2alg}. Then, any element $(X,\a,a)\in\max$ can be written in the form
\be \eqlabel{maxelement}
\left(X,\alpha,a\right)=X^\mu\bJ_\mu +\a^\mu \bPi_{\mu}+ a^\mu\bF_\mu \,.
\ee
In term of these generators, the bracket \eqref*{adESP2} leads to
\be \eqlabel{maxalg2+1}
\ba
\left[\bJ_\mu,\bJ_\nu\right]=\epsilon^\rho_{\;\mu\nu}\bJ_\rho \,,&\quad & 
\left[\bPi_\mu,\bPi_\nu\right]=\epsilon^\rho_{\;\mu\nu}\bF_\rho\,,\\
\left[\bJ_\mu,\bPi_\nu\right]=\epsilon^\rho_{\;\mu\nu}\bPi_\rho \,,&\quad &
\left[\bPi_\mu,\bF_\nu\right]=0\,,\\
\left[\bJ_\mu,\bF_\nu\right]=\epsilon^\rho_{\;\mu\nu}\bF_\rho \;, &\quad & \left[\bF_\mu,\bF_\nu\right]=0\,,
\ea
\ee
which can be put in the general form \eqref*{maxwellalg} for $D=3$ by defining the Hodge duals $\bJ_{\mu\nu}=-\epsilon^{\rho}_{\mu\nu}\bJ_\rho$, $\bF_{\mu\nu}=-\epsilon^{\rho}_{\mu\nu}\bF_\rho$. It is interesting to note that the commutation relations \eqref*{maxalg2+1} have recently been obtained in the context of non-relativistic symmetries as a particular conformal extension of the Galilean algebra \cite{Chernyavsky:2019hyp}. This extension, in turn, can be thought as the simplest example of a Hiertarinta algebra, when the role of the generators $\bPi_\mu$ and $\bF_\mu$ are interchanged \cite{Hietarinta:1975fu,Bansal:2018qyz}.

\subsubsection*{Generalizations}

The algebraic structure \eqref*{maxalg2+1} can be generalized to the
case where $K$ is an arbitrary matrix Lie group M. Thus, we can
consider $\mfk$ as a general matrix Lie algebra $\mfm$ of the
form
\be\eqlabel{genm}
\left[\bT_{a},\bT_{b}\right]=f_{\;ab}^{c}\bT_{c} \,,
\ee
where $f_{\;ab}^{c}$ are the structure constants. In that case the natural generalization of \eqref*{genmax3d} is given by $\bJ_{a} \equiv\left(\bT_{a},0,0\right)$, $\bPi_{a} \equiv\left(0,\bT_{a},0\right)$, $\bF_{a} \equiv\left(0,0,\bT_{a}\right)$, which applied to \eqref*{adESP2} leads
to
\be\eqlabel{genmaxwell}
\ba
\left[\bJ_{a},\bJ_{b}\right]=f_{\;ab}^{c}\bJ_{c}\;,&\quad & \left[\bPi_{a},\bPi_{b}\right]=f_{\;ab}^{c}\bF_{c}\,,\protect\\
\left[\bJ_{a},\bPi_{b}\right]=f_{\;ab}^{c}\bPi_{c}\;,&\quad & \left[\bPi_{a},\bF_{b}\right]=0\,,\protect\\
\left[\bJ_{a},\bF_{b}\right]=f_{\;ab}^{c}\bF_{c}\;, &\quad & \left[\bF_{a},\bF_{b}\right]=0\,.
\ea
\ee

\subsubsection*{Higher-spin extensions}

One example of such generalization is given by spin-3 extension of the Maxwell algebra found in \cite{Caroca:2017izc} as a particular expansion of the $\slt$. Now we will see that this symmetry can be obtained from \eqref{genmaxwell} when $\mfk=\slt$. A basis for $\slt$ can be defined as $\bT_a = \left\{ \bT_\mu,\,  \bTT_{\mu\nu} \right\} $, satisfying \cite{Campoleoni:2010zq}
\be\eqlabel{sltbracket}
\ba
\left[\bT_{\mu},\bT_{\nu}\right]&=&\epsilon^\rho_{\;\mu\nu}\bT_{\rho} \,,\\
\left[\bT_{\mu},\bTT_{\nu\rho}\right]&=&\epsilon^{\sigma}_{\;\mu\left( \nu\right.
}\bTT_{\left. \rho \right) \sigma} \,,\\
\left[\bTT_{\mu\nu},\bTT_{\rho\sigma}\right]&=&\varepsilon \; \eta _{ (\mu\mid \left( \rho \right.
}\;\epsilon _{\left. \sigma \right) \mid\nu) \a}  \bT^{\a}  \,,
\ea
\ee
where $\varepsilon<0$ and parentheses denote symmetrization of the enclosed indices. One can define the extended semi-direct sum algebra $\slt \oright_\ad \slt_{\rm ext}^{\rm (ab)}$, whose generators are given by $\bJ_a =\left\{ \bJ_\mu,\,  \bJJ_{\mu\nu} \right\}$, $\bPi_a = \left\{ \bPi_\mu,\,  \bPP_{\mu\nu} \right\}$, $\bF_a = \left\{ \bF_\mu,\,  \bFF_{\mu\nu} \right\}$, and have the form
\be\eqlabel{genmax3dhs}
\ba
\bJ_{\mu} \equiv \left(\bT_{\mu},0,0\right)\,, \quad
&\bPi_{\mu} \equiv \left(0,\bT_{\mu},0\right)\,, \quad
&\bF_{\mu} \equiv\left(0,0,\bT_{\mu}\right)\,,\\

\bJJ_{\mu\nu} \equiv \left(\bTT_{\mu\nu},0,0\right)\,, \quad
&\bPP_{\mu\nu} \equiv \left(0,\bTT_{\mu\nu},0\right)\,, \quad
&\bFF_{\mu\nu} \equiv\left(0,0,\bTT_{\mu\nu}\right)\,.
\ea
\ee
Using \eqref{adESP2,sltbracket} one can see that these generators satisfy the commutation relations \eqref*{maxalg2+1} plus
\be
\ba
&&\left[\bJ_{\mu},\bJJ_{\nu\rho}\right]=\epsilon^{\sigma}_{\;\mu\left( \nu\right.
}\bJJ_{\left. \rho \right) \sigma}\,,\quad
\left[\bJ_{\mu},\bPP_{\nu\rho}\right]=\epsilon^{\sigma}_{\;\mu\left( \nu\right.
}\bPP_{\left. \rho \right) \sigma}\,,\quad 
\left[\bJ_{\mu},\bFF_{\nu\rho}\right]=\epsilon^{\sigma}_{\;\mu\left( \nu\right.
}\bFF_{\left. \rho \right) \sigma}\,,\\[5pt]

&&\left[\bPi_{\mu},\bJJ_{\nu\rho}\right]=\epsilon^{\sigma}_{\;\mu\left( \nu\right.
}\bPP_{\left. \rho \right) \sigma}\,,\quad
\left[\bPi_{\mu},\bPP_{\nu\rho}\right]=\epsilon^{\sigma}_{\;\mu\left( \nu\right.
}\bFF_{\left. \rho \right) \sigma}\,,\quad 
\left[\bF_{\mu},\bJJ_{\nu\rho}\right]=\epsilon^{\sigma}_{\;\mu\left( \nu\right.
}\bFF_{\left. \rho \right) \sigma}\,,\\[5pt]

&&\left[\bJJ_{\mu\nu},\bJJ_{\rho\sigma}\right]=\varepsilon \; \eta _{ (\mu\mid \left( \rho \right.
}\;\epsilon _{\left. \sigma \right) \mid\nu) \a}  \bJ^{\a} \,,\quad 
\left[\bJJ_{\mu\nu},\bPP_{\rho\sigma}\right]=\varepsilon \; \eta _{ (\mu\mid \left( \rho \right.
}\;\epsilon _{\left. \sigma \right) \mid\nu) \a}  \bPi^{\a} \,, \\[5pt]

&&\left[\bJJ_{\mu\nu},\bFF_{\rho\sigma}\right]=\varepsilon \; \eta _{ (\mu\mid \left( \rho \right.
}\;\epsilon _{\left. \sigma \right) \mid\nu) \a} \bF^{\a} \;,\quad 
\left[\bPP_{\mu\nu},\bPP_{\rho\sigma}\right]=\varepsilon \; \eta _{ (\mu\mid \left( \rho \right.
}\;\epsilon _{\left. \sigma \right) \mid\nu) \a}  \bF^{\a} \,.

\ea
\ee
This corresponds exactly to the commutation relations of the higher-spin extension of the $\max$ algebra found in \cite{Caroca:2017izc}.  Therefore, one can define the spin-3 extension of the $\Max$ group as
\be
{\rm SL}\left(3,\mathbb{R}\right) \ltimes_\Ad \slt_{\rm ext}^{\rm (ab)} \,.
\ee
In a similar
way, more general higher-spin extensions of the Maxwell group can be defined considering the extended semi-direct product ${\rm SL}\left(N,\mathbb{R}\right) \ltimes_\Ad \sln_{\rm ext}^{\rm (ab)}$.

\subsection{Infinite-dimensional Maxwell groups in 2+1 dimensions}
\label{infmax}

In this section, we will construct two infinite-dimensional groups whose underlying structure is given by a special extended semi-direct product $K\ltimes_\Ad \mfk_{\rm \;ext}^{\rm (ab)}$ developed in Section \ref{extsdpAd}. They correspond to extensions of the loop group $L{\rm ISL}(2,\mathbb{R})$ and the $\BMS$ group.  Their Lie algebras are given by infinite-dimensional enhancements of the Maxwell algebra in 2+1 dimensions \eqref*{maxalgstructure}. These groups are interesting, as they allow for central extensions, which will be considered in section \ref{centralext}.

\subsubsection*{Maxwell loop group}

Consider the loop group $\LSl$, whose elements are the continuous maps from the unit circle $S^1$ to the group $\Sl$%
\be
\begin{array}{ll}
U: & S^{1} \longrightarrow \Sl\\ 
& \;\phi  \longmapsto U\left( \phi \right)\qquad ,\quad U\left(\phi+2\pi\right) = U\left(\phi\right)\,,
\end{array}%
\ee
and where the product of two group elements $U,V\in K$ is given by point-wise multiplication, i.e.
\be\eqlabel{prodlsl}
U\cdot V\equiv  U\left( \phi
\right) V\left( \phi \right) = \left(UV\right) \left( \phi \right)\,.
\ee
The corresponding Lie algebra $\Lsl$ is the algebra of continuous maps from $S^{1}$ to $\sl$. As functions on the unit circle can be expanded in Fourier series, the elements of $\Lsl$ can be written as $X=X^\mu(\phi)\bt_\mu=\sum_{m=-\infty}^{\infty} X_m^\mu \bt_\mu^m $, where the generators of the loop algebra $\Lsl$  can be written as
\be\eqlabel{genlsl}
\bt^m_\mu = e^{im\phi} \bt_\mu\,,
\ee
and satisfy the commutation relations
\be\eqlabel{lsl2alg}
\left[\bt^m_{\mu},\bt^n_{\nu}\right]=\epsilon_{\;\mu\nu}^{\rho}\bt^{m+n}_{\rho}\;.
\ee

Similarly to what happens with the Maxwell algebra, the dual space $\Lsl^*$ is isomorphic to $\Lsl$. Its elements can be written as $J=J_\mu(\phi)\bt^\mu=\sum_{m=-\infty}^{\infty} J_\mu^m \bt^\mu_m$, where the dual basis is constructed out of the invariant $\sl$ metric $\eta_{\mu\nu}$ as $\bt^\mu _m=e^{im\phi}\eta^{\mu\nu} \bt_\nu$. The corresponding pairing is then given by
\be\eqlabel{pairinglsl}
\left\langle J, X\right\rangle_{L\,\sl}=\frac{1}{2\pi}\int_{0}^{2\pi } \hspace{-5pt} d\phi  \, \left\langle J, X \right\rangle_{\sl} = \frac{1}{2}  \sum_{m=-\infty}^{\infty}  J_\mu ^{m} X^\mu _{-m} \,.
\ee
Therefore, the adjoint and the coadjoint representations of $\LSl$ are given by expressions similar to \eqref*{adcoadsl}, 
\be\eqlabel{adcoadsl2}
\Ad_{U}X=U X U^{-1}\; ,\quad
\Ad_{U}^{*}J=U J U^{-1}\;, \quad
\ad_{X}Y= \left[X,Y\right] \; , \quad
\ad_{X}^{*}J=\left[X,J\right] \,,
\ee
where now the group elements as well as the elements of the corresponding Lie algebra and its dual space are functions on $S^1$.

We can define the $\Max$ loop group as the extended semi-direct product \eqref*{esdpad}, where $K=\LSl$. i.e.
\be\eqlabel{LMax}
\LMax = \LSl\ltimes_{{\rm Ad}}\Lsl_{\rm ext}^{\rm (ab)} \,.
\ee
Here $\Lsl_{\rm ext} ^{\rm (ab)}$ is given by $\Lsl^{\rm (ab)} \times \Lsl ^{\rm (ab)}$ as a set, and its product is defined by \eqref{prodslext}, with \eqref*{sl2alg} replaced by \eqref*{lsl2alg}.

Following the construction developed in Section \ref{extsdpAd}, the Maxwell loop algebra in 2+1 dimensions, $\Lmax$, is given by the semi-direct sum $\Lsl\oright_\ad \Lsl_{\rm \,ext}^{\rm (ab)}$. Its elements can be written in a form analogue to \eqref*{maxelement} with $\phi$-dependent components, which can be expanded in Fourier modes as
\be
\left(X,\alpha,a\right)=X^\mu(\phi)\bJ_\mu +\a^\mu(\phi) \bPi_{\mu}+ a^\mu(\phi) \bF_\mu=\sum_{m=-\infty}^{\infty}\left(X^\mu_m \bJ^m_\mu +\a^\mu_m \bPi^m_{\mu}+ a^\mu_m\bF^m_\mu\right)\,,
\ee
where $
\bJ^m_{\mu}= e^{im\phi} \bJ_\mu$, $
\bPi^m_{\mu}= e^{im\phi} \bPi_\mu$, and $
\bF^m_{\mu}= e^{im\phi} \bF_\mu$ form a base for the Maxwell loop algebra. Using the definitions \eqref*{genmax3d,genlsl}, these generators can be written as
\be\eqlabel{genmax3d2}
\bJ^m_{\mu} \equiv\left(\bt^m_{\mu},0,0\right)\,, \quad
\bPi^m_{\mu} \equiv\left(0,\bt^m_{\mu},0\right)\,, \quad
\bF^m_{\mu} \equiv\left(0,0,\bt^m_{\mu}\right)\,,
\ee
and its commutation relations follow from \eqref*{adESP2} and \eqref*{lsl2alg}, which leads to
\be\eqlabel{lmaxalg}
\ba
\left[\bJ^m_{\mu},\bJ^n_{\nu}\right]=\epsilon_{\;\mu\nu}^{\rho}\bJ^{m+n}_{\rho}\;,&\quad & \left[\bPi^m_{\mu},\bPi^n_{\nu}\right]=\epsilon_{\;\mu\nu}^{\rho}\bF^{m+n}_{\rho}\,, \\
\left[\bJ^m_{\mu},\bPi^n_{\nu}\right]=\epsilon_{\;\mu\nu}^{\rho}\bPi^{m+n}_{\rho}\;,&\quad & \left[\bPi^m_{\mu},\bF^n_{\nu}\right]=0\,, \\
\left[\bJ^m_{\mu},\bF^n_{\nu}\right]=\epsilon_{\;\mu\nu}^{\rho}\bF^{m+n}_{\rho}\;, &\quad & \left[\bF_{\mu},\bF_{\nu}\right]=0\,.
\ea
\ee
The elements of the dual space are naturally constructed as triples $(J,\Pi,F)$, where $J\in\Lsl^*$ and $\Pi,F\in\Lsl^{\rm (ab)*}$. The pairing between the $L\max$ algebra and its dual follows directly from \eqref*{pairingesdp2,pairinglsl}, while the adjoint and coadjoint representations (together with their infinitesimal forms) can be constructed in a straightforward way. The geometric action \eqref*{espga2} in this case is given by a sigma model of the form 
\be\eqlabel{actionLMax}
I_{\LMax}=\frac{1}{2\pi} \int_0^{2\pi} \hspace{-5pt} d\phi \int dt \, {\rm Tr} \left[ J^0 \, U^{-1}\dot U
+
U \Pi^0 U^{-1} \, \dot\alpha 
+
U F^0 U^{-1} \, \left(\dot a+ \frac{1}{2} \left[\alpha,\dot\alpha\right]\right)\right]\,.
\ee

This construction can be easily generalized to more general loop groups with Maxwell structure by considering an extended semi-direct product \eqref*{esdpad} where $K$ is the loop group $LM$ associated to a matrix Lie group $M$
\be\eqlabel{loopmatrix}
LM\ltimes_\Ad L\mathfrak{m}_{\rm ext}^{\rm (ab)}\,.
\ee
This will lead to algebras of the form $L\mfm\oright_\ad L\mfm_{\rm ext}^{\rm (ab)}$ whose generators are given by 
\be\eqlabel{basekmgen}
\bJ^m_{a} \equiv\left(e^{im\phi}\bT_{a},0,0\right)\;,\quad \bPi^m_{a} \equiv\left(0,e^{im\phi}\bT_{a},0\right) \;,\quad \bF^m_{a} \equiv\left(0,0,e^{im\phi}\bT_{a}\right)\,,
\ee
and $\bT_a$ stands for the generators of $\mfm$ defined in \eqref{genm}. The base \eqref*{basekmgen} will automatically satisfy an algebra of the form \eqref*{lmaxalg} with $\epsilon^{\rho}_{\;\mu\nu}\rightarrow f^{c}_{\;ab}$. The geometric action in this case takes a form analogue to \eqref{actionLMax} where the pairing \eqref*{pairinglsl} is to be replaced by integration on $S^1$ of the corresponding the Killing form on $M$.

\subsubsection*{Extended $\BMS$ group}

The second infinite-dimensional Maxwell group we will be interested in is based on the orientation-preserving diffeomorphism group of the unit circle $\Diff$. In this case, an element $U\in\Diff $ is a reparametrization of $S^1$:
\be\eqlabel{diffcircle}
\begin{array}{ll}
U: & S^{1}\longrightarrow S^{1} \\
& \phi \longmapsto U\left( \phi \right)\,,\quad U\left(\phi+2\pi\right) = U\left(\phi\right)+2\pi\,,\quad U^{\prime}(\phi)>0\,,
\end{array}%
\ee
where prime denotes derivative with respect to $\phi$ and the group operation is given by function composition
\be\eqlabel{grouplawdiff}
U\cdot V = U\circ V\,.
\ee
The corresponding Lie algebra is the algebra of vector fields on the circle $\Vect$, whose elements will be denoted by $X=X(\phi)\d_\phi$. The elements of the dual space $\Vect^*$ correspond to quadratic differentials on $S^{1}$, $J=J\left( \phi \right) \left( d\phi \right)^{2} $, and the paring is given by
\be\eqlabel{pairingwitt}
\left\langle J,X\right\rangle_{\Vect} =\int_{0}^{2\pi }d\phi J\left( \phi \right) X\left( \phi
\right) \,.
\ee
The adjoint and the coadjoint representations of $\Vect$ are defined by the action of diffeomorphisms on elements of $\Vect$ and $\Vect^*$, respectively \cite{Barnich:2017jgw,Oblak:2017ect}
\be\eqlabel{adcoaddiff}
  {\rm Ad}_{U}\,X = 
  \frac{   X\circ U^{-1}\left(\phi\right)   }{    \left(U^{-1}\right)' \! \left(\phi\right)   }  \partial_{\phi}\;,\quad
{\rm Ad}_{U}^{\ast}\,J =  \left[{\left(U^{-1}\right)^{\prime}  \left(\phi\right)}\right]^{2}
  \; J\circ U^{-1}\left(\phi\right)d\phi^{2}\,,
\ee
while the infinitesimal adjoint and coadjoint representations of $\Vect$ read 
\be\eqlabel{infadcoaddiff}
\ba
\ad_{X}Y=-\Big( X(\phi)Y^\prime (\phi)-X^\prime(\phi)
Y(\phi)\Big)
\d_\phi \;,\\
\ad_{X}^{\ast }J=-\Big( J^{\prime }(\phi )X(\phi )+2J(\phi )X^{\prime }(\phi
)\Big) \left(d\phi \right)^{2}\,. 
\ea
\ee
Defining the generators \be\eqlabel{wittgen}\bell_m=-i e^{i m \phi }\d_\phi\,,\ee  and expanding the elements in $\Vect$ in Fourier modes as $X=\frac{1}{2\pi}\sum_{m=-\infty}^\infty X_m \bell_m$, the bracket defined in \eqref{infadcoaddiff} leads to the Witt algebra
\be\eqlabel{witt}
\left[ \bell_m , \bell_n \right] = \left(m-n\right) \bell_{m+n}\,.
\ee
Now we define the \emph{Maxwell-like extension} of the $\BMS$ group as the extended semi-direct product
\be\eqlabel{MBMS}
\MBMS = \Diff \ltimes_\Ad \Vect^{{\rm (ab)}}_{\rm ext}\,,
\ee
where $\Vect^{{\rm (ab)}}_{\rm ext}$ is the set $\Vect^{{\rm (ab)}}\times \Vect^{{\rm (ab)}}$ endowed with an extended sum given by \eqref*{exttranslations,xprodad}. The elements of $\MBMS$ are given by triplets $\left(U,\a,a\right)$ with $U\in\Diff$ and $\a,a\in\Vect$ and its group law can be read off from  \eqref*{groupop2}, using \eqref{grouplawdiff,adcoaddiff,infadcoaddiff}. Similarly, the adjoint and the coadjoint representations follow from \eqref*{AdESP2,adcoad2}, and the geometric action \eqref*{espga2} can be evaluated to give
\be\eqlabel{actionMBMS}
I_{\MBMS}=\int_0^{2\pi}  \hspace{-5pt} d\phi \int dt \,  \frac{1}{U^{\prime}} \left[  J^0 \dot U + \Pi^0 \;\; \dot \a\!\circ\! U
 + F^0 \;\left( \dot a  + \frac{1}{2} \left( \a^{\prime} \dot \a- \a\dot \a^{\prime} \right) \right)\!\circ\! U  \right]\,,
\ee
where here $\a$ and $a$ denote their components $\a(\phi)$ and $a(\phi)$, and similarly for the elements $\left(J_0,\Pi_0,F_0 \right)$ in the dual space $\Vect^*$.

The corresponding Lie algebra,
\be
\mbms=\Vect\oright_\ad \Vect^{\rm (ab)}_{\rm ext} \,,
\ee
defines an extension of the $\bms$ algebra with Maxwell structure. The Lie bracket has the form \eqref*{adESP2}, which can be written in terms of the generators
\be\eqlabel{gensextbms}
\bJJ_m \equiv\left(\bell_m,0,0\right)\,, \quad
\bPP_m \equiv\left(0,\bell_m,0\right)\,, \quad
\bFF_m \equiv\left(0,0,\bell_m \right)\,,
\ee
and yields the following commutation relations
\be
\ba
\left[\bJJ_{m},\bJJ_{n}\right]  =\left(m-n\right)\bJJ_{m+n}\,,&\quad &\left[\bPP_{m},\bPP_{n}\right] =\left(m-n\right)\bFF_{m+n}\,,\\
\left[\bJJ_{m},\bPP_{n}\right] =\left(m-n\right)\bPP_{m+n}\,,&\quad&\left[\bPP_{m},\bFF_{n}\right] = 0\,,\\
\left[\bJJ_{m},\bFF_{n}\right] =\left(m-n\right)\bFF_{m+n}\,,&\quad &\left[\bFF_{m},\bFF_{n}\right] = 0\,.
\ea
\ee

\section{Centrally extended Maxwell-like groups in 2+1 dimensions}
 \label{centralext} 

The special structure the extended semi-direct product $K\ltimes_\Ad \mfk_{\rm ext}^{\rm (ab)}$
defined in Section \ref{extsdpAd} allows one to introduce a central extension
in a straightforward way, provided a central extension of $K$ is known. This can be achieved by generalizing the construction of centrally extended semi-direct products \cite{Barnich:2014kra,Barnich:2015uva}.

A central extension $\hat{K}$ of a group $K$ consists in a direct
product $\widehat{K}=K\times\mathbb{R}$ with elements $\left(U,m\right)$,
where $U\in K$ and $m\in\mathbb{R}$, and group operation
\begin{equation}\eqlabel{productceg}
\left(U,m\right)\left(V,n\right)=\left(U\cdot V,m+n+\mathcal{C}\left(U,V\right)\right)\,,
\end{equation}
where $\mathcal{C}\left(U,V\right)$ defines a two-cocycle on $K$, i.e.
\be
\mathcal{C}(U\cdot V,W)+C(U,V)=\mathcal{C}(U,V\cdot W)+\mathcal{C}(V,W)\,.
\ee
The corresponding centrally extended algebra $\hat{\mathfrak{k}}$
has elements $\left(X,m\right)$ and its Lie bracket has the form
\begin{equation}
\left[\left(X,m\right),\left(Y,n\right)\right]=\left(\left[X,Y\right],  \mathfrak{c}\left(X,Y\right)\right)\,,\eqlabel{bracketce}
\end{equation}
where $\mathfrak{c}$ is a Lie algebra two-cocycle, $\mathfrak{c}\in H^{2}\left(\mathfrak{k},\text{\ensuremath{\mathbb{R}}}\right)$,
given by
\be\eqlabel{algcocycle}
\mathfrak{c}(X,Y)=\left.\frac{d}{d\lambda_1}\frac{d}{d\lambda_2}\left[\mathcal{C}(e^{\lambda_1 X},e^{\lambda_2 Y})-\mathcal{C}(e^{\lambda_2 Y},e^{\lambda_1 X})\right]\right\vert _{\lambda_1=\lambda_2=0}\,.
\ee

\subsection{Central extensions of extended semi-direct products}

Let us consider a special extended semi-direct product $H=K\ltimes_\Ad \mfk^{\rm (ab)}_{\rm \,ext}$ as defined in \eqref*{esdpad}. If a non-trivial central extended group $\hat{K}$ can be constructed,
the central extension of $H$ can be
defined as
\be\eqlabel{extH}
\hat{H}=\hat{K}\ltimes_\Ad \hat{\mfk}^{\rm (ab)}_{\rm \,ext}\,.
\ee
Its elements are sextuplets $\left(U,m_{1};\a,m_{2}; a,m_{3}\right)$ where $m_1$, $m_2$ and $m_3$ are central terms,
and the group operation can be obtained from \eqref{groupop1} by implementing the following generalization
\begin{equation}\eqlabel{prescription}
U\rightarrow\left(U,m_{1}\right)\,,\quad
\a\rightarrow\left(\a,m_{2}\right)\,,\quad
a \rightarrow\left(a ,m_{3}\right)\,.
\end{equation}
This yields to the following product law:
\begin{equation}\eqlabel{productceg2}
\left(U,m_{1};\a,m_{2};a,m_{3}\right)\bullet\left(V,n_{1};\b,n_{2};b ,n_{3}\right)=\left(W,q_{1};\omega ,q_{2};w,q_{3}\right)\,,
\end{equation}
where
\begin{equation}\eqlabel{productceg3}
\begin{array}{lcl}
\left(W, q_1\right) & = & \left(U,m_1\right)\left(V,n_1\right) \,,\\
\left(\omega,q_2\right) & = & \left(\a,m_2\right)+\Ad_{\left(U,m_1\right)}\left(\b,n_2\right)  \,, \\
\left( w,q_3\right) & = & \left(a,m_3\right)+\Ad_{\left(U,m_1\right)}\left(b,n_3\right)+\frac{1}{2}\ad_{\left(\a,m_2\right)}\Ad_{\left(U,m_1\right)}\left(\b,n_2\right)\,.
\end{array}
\end{equation}
Here, the expressions $\Ad_{\left(U,m_1\right)}$
and $\ad_{\left(\a,m_2\right)}$
denote the adjoint representation of the centrally extended group $\hat{K}$ and the centrally extended algebra $\hat\mfk\,$, respectively. The former can be obtained in the usual way by differentiating the conjugation operation in $\hat{K}$ with the product law \eqref*{productceg}. This  yields \cite{khesin2008geometry}
\be
\Ad_{\left(U,m\right)}\left(X,n\right)=\left(\Ad_U X, n-\left\langle \cS\left(U\right),X\right\rangle \right)\,,
\ee
where $\cS$ is the Souriau cocyle on $K$, which
is defined in terms of the two-cocycle $\cC$ as
\be\eqlabel{souriau}
\left\langle \cS\left(U\right),X\right\rangle =-\left.\frac{d}{d\lambda}\left[\cC(U,e^{\lambda X}U^{-1})+\cC(e^{\lambda X},U^{-1})\right]\right\vert _{\lambda=0} \,.
\ee
The later is the infinitesimal form of ${\rm {\rm Ad}}_{\text{\ensuremath{\left(U,m\right)}}}$, which leads to the bracket \eqref*{bracketce}, and has the form
\be
\ad_{\left(A,m\right)}\left(B,n\right)=\left({\rm ad}_{A}B,-\left\langle \mathfrak{s}\left(A\right),B\right\rangle \right) \,,
\ee
where $\mathfrak{s}$ corresponds to the differential of $\cS$ evaluated at the identity, and it is related with the two-cocycle $\mathfrak{c}$ defined in \eqref*{algcocycle} by
\be\eqlabel{algcocycle2}
\mathfrak{c}(A,B)=-\left\langle \mathfrak{s}\left(A\right),B\right\rangle  \,.
\ee
Replacing these expressions in \eqref{productceg3} leads to
\be
\begin{array}{lcl}
W=U\cdot V\,, & \quad & q_{1}=m_{1}+n_{1}+\cC\left(U,V\right)\,,\\[5pt]
\omega=\a + \Ad_{U}\b\,, & \quad & q_{2}=m_{2}+n_{2}-\left\langle \cS(U),\b\right\rangle \,,\\[5pt]
w=a+\Ad_U b+\frac{1}{2}\ad_\a\Ad_U\b\,, & \quad & q_{3}=m_{3}+n_{3}-\left\langle \mathcal{S}(U),b \right\rangle -\frac{1}{2}\left\langle \mfs( \a),\Ad_U \b \right\rangle \,,
\end{array}\eqlabel{productceg4}
\ee
which determines the group operation \eqref*{productceg2}.

The corresponding centrally extended algebra is given by $\hat{\mathfrak{h}}=\hat{\mfk}\oright_\ad \hat{\mfk}^{\rm (ab)}_{\rm ext}$. Denoting its elements by $(X,m_1,\a,m_2,a,m_3)$, the associated bracket can be obtained by computing the adjoint representation of $\hat{H}$, which follows from the definition \eqref*{AdESP1} and the prescription \eqref*{prescription},
\be\eqlabel{Adcentrallyext0}
\Ad_{\left(U,m_{1};\b,m_{2};b,m_{3}\right)}\left(X,n_{1};\a,n_{2};a,n_{3}\right)=\left(\tilde{X},\tilde{n}_{1};\tilde{\a},\tilde{n}_{2};\tilde{a},\tilde{n}_{3}\right) \,,
\ee
where
\be\eqlabel{Adcentrallyext}
\ba
\tilde{X} & = & \Ad_ U X  \,,\\[5pt]
\tilde{n}_{1} & = & n_{1}-\left\langle \cS\left(U\right),X\right\rangle \,, \\[5pt]
\tilde{\a} & = &\Ad_ U\a-\ad_{\Ad_ U X}\b \,,\\[5pt]
\tilde{n}_{2} & = & n_{2}-\left\langle \cS(U),\a \right\rangle +\left\langle \mfs(\Ad_ U X),\b\right\rangle \,, \\[5pt]
\tilde{a} & = & \Ad_ U a-\ad_{\Ad_ U X}b+\ad_\b\left(\Ad_ U \a+\dfrac{1}{2}\ad_\b \Ad_ U X\right) \,, \\[5pt]
\tilde{n}_{3} & = & n_{3}-\left\langle \cS(U),a\right\rangle +\left\langle \mfs(\Ad_ U X),b\right\rangle -\left\langle \mfs(\b),\Ad_ U \a+\dfrac{1}{2}\ad_\b \Ad_U X\right\rangle \,.
\ea
\ee
The infinitesimal limit of the expression \eqref*{Adcentrallyext} leads to the following bracket
\be\eqlabel{ceesdpalg}
\ba
\left[\left(X,m_{1};\a,m_{2}; a,m_{3}\right),\left(Y,n_{1};\b,n_{2}; b ,n_{3}\right)\right]=\\[6pt]
\qquad\qquad=\left(\left[X,Y\right],f_{1};\left[X,\b\right]-\left[Y,\a\right],f_{2};\left[X,b\right]-\left[Y,a\right]+\left[\a,\b\right],f_{3}\right) \,,
\ea
\ee
where the central elements are written in terms of the Lie algebra cocycle \eqref*{algcocycle} as
\be
\ba
f_{1} & = & \mathfrak{c}\left(X,Y\right) \,, \\
f_{2} & = & \mathfrak{c}\left(X,\b\right)-\mathfrak{c}\left(Y,\a\right) \,,\\
f_{3} & = & \mathfrak{c}\left(X,b\right)-\mathfrak{c}\left(Y,a\right)+\mathfrak{c}\left(\a,\b\right) \,.
\ea
\ee
In the following, we will see how to apply this construction to the infinite-dimensional Maxwell-like groups in 2+1 dimensions studied in Section \ref{infmax}. This will lead to novel centrally extended groups, whose associated infinite-dimensional algebras have been recently  discussed in \cite{Caroca:2017onr,Concha:2018zeb} from the point of view of Lie algebra expansions and asymptotic symmetries in three-dimensional gravity theories.

\subsection{Maxwell-Kac-Moody group}

Let us construct now the central extension of the Maxwell loop group in 2+1 dimensions defined in \eqref*{LMax}. Following \eqref*{extH}, first we need to define the central extension of the loop group $\LSl$. This corresponds to the Kac-Moody group $\hatLSl$, defined by the following two-cocycle
\be\eqlabel{comk}
\cC\left(U,V\right)=\frac{1}{4\pi}\int_{D^1}{\rm Tr}\left[U^{-1}\bar{\rm  d}U\, \bar{\rm d}V\,V^{-1} \right] \,,
\ee
where the exterior $\bar {\rm d}$ is defined on the unit disc $D^1$. This two-cocyle extends the product \eqref*{prodlsl} in $\LSl$ in the form \eqref*{productceg}. The definition \eqref*{bracketce} leads to the Kac-Moody algebra $\hatLsl$, where the two-cocycle \eqref*{algcocycle}  reads 
\be
\mathfrak{c}(X,Y)=\frac{1}{2\pi}\int_{0}^{2\pi} \hspace{-5pt} d\phi  \,  X(\phi)\,Y^{\prime}(\phi) \,.
\ee
The corresponding Souriau cocyle and its differential can be computed using \eqref*{souriau} and have the form
\be\eqlabel{souriaulsl}
\cS\left( U\right) =\frac{1}{2\pi}U^\prime U ^{-1}\;,\quad \mfs\left( X\right) =\frac{1}{2\pi}X^\prime (\phi) \,.
\ee
Now we define the \emph{Maxwell-Kac-Moody  group} in 2+1 dimensions as the extended semi-direct product
\be\eqlabel{hatLMax}
\hatLMax=\hatLSl \ltimes_\Ad \hatLsl_{\rm ext}^{\rm (ab)} \,,
\ee
whose product law has the form given in \eqref{productceg3,productceg4} and its written in terms of the adjoint representation of $\Lsl$, given in \eqref*{adcoadsl2}, and the cocycles \eqref*{comk,souriaulsl} previously defined. The \emph{Maxwell-Kac-Moody algebra} in 2+1 dimensions is therefore given by
\be
\hatLmax= \hatLsl \oright_\ad \hatLsl_{\rm ext}^{\rm (ab)} \,.
\ee
The adjoint representation of the $\hatLMax$ group as well as the commutation relations of the $\hatLmax$ algebra can be directly read off from \eqref*{Adcentrallyext,ceesdpalg} for $\mfk=\Lsl$. The commutation relations can be expressed in terms of a base of the form
\be
\ba
\bJ^m_{\mu} &=\left(e^{im\phi}\bt_{\mu},0;0,0;0,0\right)\;,\quad 
\bk_1 &= \left(0,1;0,0;0,0\right)\,,\quad  \\

\bPi^m_{\mu} &=\left(0,0;e^{im\phi}\bt_{\mu},0;0,0\right) \;,\quad
\bk_2& =\left(0,0;0,1;0,0\right)\,,\quad \\

\bF^m_{\mu} &=\left(0,0;0,0;e^{im\phi}\bt_{\mu},0\right) \;,\quad
\bk_3 &=\left(0,0;0,0;0,1\right)\,,
\ea
\ee
where $\bt_\mu$ is the generator of the $\sl$ algebra \eqref*{lsl2alg}. In terms of these generators, \eqref{ceesdpalg} leads to a centrally extended version of the algebra \eqref*{lmaxalg},
\be\eqlabel{maxkmalg}
\ba
\left[\bJ^m_{\mu},\bJ^n_{\nu}\right]=\epsilon_{\;\mu\nu}^{\rho}\bJ^{m+n}_{\rho}+\bk_{1}m\,\delta_{\mu\nu}\delta^{m,-n}\,,&\quad & \left[\bPi^m_{\mu},\bPi^n_{\nu}\right]=\epsilon_{\;\mu\nu}^{\rho}\bF^{m+n}_{\rho}+\bk_{3}m\,\delta_{\mu\nu}\delta^{m,-n}\,, \\[5pt]
\left[\bJ^m_{\mu},\bPi^n_{\nu}\right]=\epsilon_{\;\mu\nu}^{\rho}\bPi^{m+n}_{\rho}+\bk_{2}m\,\delta_{\mu\nu}\delta^{m,-n}\,,&\quad & \left[\bPi^m_{\mu},\bF^n_{\nu}\right]=0\,, \\[5pt]
\left[\bJ^m_{\mu},\bF^n_{\nu}\right]=\epsilon_{\;\mu\nu}^{\rho}\bF^{m+n}_{\rho}+\bk_{3}m\,\delta_{\mu\nu}\delta^{m,-n}\,, &\quad & \left[\bF_{\mu},\bF_{\nu}\right]=0\,.
\ea
\ee
This structure can be generalized to more general Kac-Moody groups with Maxwell structure by centrally extending groups of the form \eqref*{loopmatrix}. This can be done by first extending the loop group $LM$ by means of a two-cocycle of the form \eqref*{comk} where the trace over $\sl$ elements is to be replaced by the Killing form on $\mfm$. The resulting Lie algebra will have the form $\widehat{L\mfm}\oright_\ad \widehat{L\mfm}_{\rm ext}^{\rm (ab)}$ and its commutation relations will have the structure \eqref*{maxkmalg} with more general structure constants, defining a Kac-Moody version of \eqref*{genmaxwell}. This type of algebras have been found in \cite{Caroca:2017onr} as expansions of a Kac-Moody algebra $\widehat{L \mfm}$.

\subsection{Maxwell-like extension of the $\hatBMS$ group}

In this section, we will define and extension of the $\hatBMS$ group using the group structure \eqref*{extH}. The starting point is the Virasoro group $\hatDiff$, which is the central extension of the diffeomorphism group of the circle defined in \eqref{diffcircle,grouplawdiff}. It is  defined by the Thurston-Bott cocycle \cite{khesin2008geometry},
\be\eqlabel{btcc}
\cC\left(U,V\right)=-\frac{1}{48\pi}\int_{0}^{2\pi}  \hspace{-5pt} d\phi \, {\rm Log}\left[U^{\prime}\circ V\right]{\rm Log}\left[V^{\prime}\right]^{\prime} \,,
\ee
which extends the function composition product \eqref*{grouplawdiff} in the form \eqref*{productceg}. The corresponding Souriau cocycle is given by  the Schwarzian derivative
\begin{equation}\eqlabel{souriauvir}
\cS\left( U\right) =\frac{U^{^{\prime \prime \prime }}}{U^{^{\prime }}}-\frac{3%
}{2}\left( \frac{U^{^{\prime \prime }}}{U^{^{\prime }}}\right) ^{2} 	\,,
\end{equation}
and the bracket of the Virasoro algebra $\hatVect$ follows from \eqref{bracketce} by either using the definition \eqref*{algcocycle} or \eqref*{algcocycle2}, where
\be\eqlabel{infsouriauvir}
\mathfrak{s}(X)=\frac{1}{24\pi}X^{\prime\prime\prime}(\phi)\partial_\phi\,, \quad \mathfrak{c}(X,Y)=\frac{1}{24\pi}\int_{0}^{2\pi} \hspace{-5pt} d\phi \, X(\phi)\,Y^{\prime\prime\prime}(\phi)\,.
\ee
Now we consider the following extended semi-direct product based on the Virasoro group 
\be\eqlabel{MhatBMS}
\MhatBMS=\hatDiff \ltimes_\Ad \hatVect_{\rm ext}^{\rm (ab)} \,.
\ee
This is the natural \emph{Maxwell-like extension} of the centrally extended ${\rm BMS}$ group in 2+1 dimensions, which can be defined as the semi-direct product of the Virasoro group and its Lie algebra \cite{Barnich:2014kra}. These definitions, together with the adjoint representation of the $\Diff$ group given in \eqref*{adcoaddiff,infadcoaddiff}, define the group operation of the $\MhatBMS$ group through \eqref{productceg2}. The Lie algebra
\be\eqlabel{mbms3algebra}
\mhatbms=\hatVect \oright_\ad \hatVect_{\rm ext}^{\rm (ab)} \
\ee
defines a Maxwell extension of the $\hatbms$ algebra, whose bracket is defined by \eqref{ceesdpalg}. Considering a base of the form
\be\eqlabel{basembms3}
\ba
\bJJ_m &=\left(\bell_m,0;0,0;0,0\right)\;,\quad 
\bc_1 &= \left(0,1;0,0;0,0\right)\,,\quad  \\

\bPP_m &=\left(0,0;\bell_m,0;0,0\right) \;,\quad
\bc_2& =\left(0,0;0,1;0,0\right)\,,\quad \\

\bFF_m &=\left(0,0;0,0;\bell_m,0\right) \;,\quad
\bc_3 &=\left(0,0;0,0;0,1\right)\,,
\ea
\ee
where $\bell_m$ stands for the generator of the Witt algebra \eqref*{witt}, leads to
\be\eqlabel{mbms3}
\ba
\left[\bJJ_{m},\bJJ_{n}\right] =\left(m-n\right)\bJJ_{m+n}+\dfrac{\bc_{1}}{12}\,m^{3}\delta_{m,-n}\;,&\quad & \left[\bPP_{m},\bPP_{n}\right] =\left(m-n\right)\bFF_{m+n}+\dfrac{\bc_{3}}{12}\,m^{3}\delta_{m,-n}\,, \\[6pt]
\left[\bJJ_{m},\bPP_{n}\right] =\left(m-n\right)\bPP_{m+n}+\dfrac{\bc_{2}}{12}\,m^{3}\delta_{m,-n}\;,&\quad & \left[\bPP_m,\bFF_n\right]=0\,, \\[6pt]
\left[\bJJ_{m},\bFF_{n}\right] =\left(m-n\right)\bFF_{m+n}+\dfrac{\bc_{3}}{12}\,m^{3}\delta_{m,-n}\;, &\quad & \left[\bFF_{\mu},\bFF_{\nu}\right]=0\,,
\ea
\ee
where we have used \eqref*{wittgen,infsouriauvir}. These commutation relations have been found in \cite{Caroca:2017onr} as an expansion of the Virasoro algebra. Furthermore, they can be obtained from the Maxwell-Kac-Moody algebra \eqref*{maxkmalg} by means of a generalized Sugawara construction. On the other hand, the algebra \eqref*{mbms3} has been found in \cite{Concha:2018zeb} as the asymptotic symmetry of a three-dimensional Chern-Simons
gravity theory invariant under the $\max$ algebra.

\subsection{Wess-Zumino terms}
\label{wzwterms}

In this section, we construct the geometric actions for the $\hatLMax$ group and the $\MhatBMS$ group studied in the previous section, by generalizing the construction shown in Section \ref{extsdpAd} to centrally extended groups of the form \eqref*{extH}. In order to do this, we need to define the dual space $\hat{\mfh}^*$. Its elements will be denoted by sextuplets of the form $\left(J,c_1,\Pi,c_2,F,c_3\right)$, and the pairing between $\hat \mfh$ and its dual space can be defined as the natural extension of \eqref{pairingesdp2}, i.e.
\be\eqlabel{pairingcen}
\left\langle \left( J,c_{1};\Pi,c_{2};F,c_{3} \right),\left( X,m_{1};\a,m_{2};a,m_{3}\right)\right\rangle_{\hat{\mathfrak{h}}}=\left\langle \left(X,\Pi,J\right),\left(X,\a,a\right)\right\rangle_\mathfrak{h} + c_1 m_1 + c_2 m_2 + c_3 m_3  \,.
\ee
The coadjoint representation of $\hat{H}$ can be defined by using the prescription
\be\eqlabel{prescriptiondual}
J\rightarrow\left(J,c_{1}\right)\;,\quad \Pi\rightarrow\left(\Pi,c_{2}\right)\;,\quad F\rightarrow\left(F,c_{3}\right) \,,
\ee
together with \eqref*{prescription} in the general expression for special extended semi-direct products \eqref*{adcoad2}. This leads to 
\be\eqlabel{coadceesdp}
\Ad_{\left(U,m_{1};\a,m_{2};a,m_{3}\right)}^{*}\left(J,c_{1};\Pi,c_{2};F,c_{3}\right)=\left(\tilde{J}, c_{1};\tilde{\Pi}, c_{2};\tilde{F}, c_{3}\right) \,,
\ee
where
\be
\ba
\tilde{J} = \Ad_{U}^{*}J-c_{1}\cS\left(U^{-1}\right)+\ad_{\a}^{*}\Big(\Ad_{U}^{*}\Pi-c_{2}\cS\left(U^{-1}\right)\Big)+c_{2}\mathfrak{s}\left(\a\right)\\[6pt]
 +\,\ad_{a}^{*}\Big(\Ad_{U}^{*}F-c_{3}\cS\left(U^{-1}\right)\Big)+c_{3}\mathfrak{s}\left(a\right)+\frac{1}{2}\ad_{\a}^{*}\Big[\ad_{\a}^{*}\Big(\Ad_{U}^{*}F-c_{3}\cS\left(U^{-1}\right)\Big)+c_{3}\mathfrak{s}\left(\a\right)\Big] \,, \\[8pt]
\tilde{\Pi} =\Ad_{U}^{*}\Pi-c_{2}\cS\left(U^{-1}\right)+\ad_{\a}^{*}\left(\Ad_{U}^{*}F-c_{3}\cS\left(U^{-1}\right)\right)+c_{3}\mathfrak{s}\left(\a\right) \,,\\[7pt]
\tilde{F} =\Ad_{U}^{*}F-c_{3}\cS\left(U^{-1}\right)\,. 
\ea
\ee
Similarly, the infinitesimal coadjoint action of $\hat \mfh$ on its dual space can be generalized from \eqref{infcoad}. Given a fixed element $ \left( J_0,c_{1};\Pi_0,c_{2};F_0,c_{3} \right)\in\hat{\mathfrak{h}}$, the expression \eqref*{coadceesdp} can be used to define the coadjoint orbits $O_{ \left( J_0,c_{1};\Pi_0,c_{2};F_0,c_{3} \right)}$ of $\hat{H}$ according to \eqref*{orbit1}. Then, the expression of the corresponding geometric action follows from the direct generalization of \eqref{geomaction},
\be\eqlabel{geomactionce}
I_{\hat H}=\int_\Gamma \left\langle \left(J_0,c_1 ; \Pi_0,c_2 ;F_0,c_3 \right) , \left(\Theta_{U},\Theta_{1};\Theta_{\a},\Theta_{2};\Theta_{a},\Theta_{3}\right) \right\rangle_{\hat{\mathfrak{h}}} \,,
\ee
where $\left(\Theta_{U},\Theta_{1};\Theta_{\a},\Theta_{2};\Theta_{a},\Theta_{3}\right)$ stands for the left-invariant Maurer-Cartan form on $\hat{H}$ and satisfies
\be
d\left(\Theta_{U},\Theta_{1};\Theta_{\a},\Theta_{2};\Theta_{a},\Theta_{3}\right)=-\frac{1}{2}\left[ \left(\Theta_{U},\Theta_{1};\Theta_{\a},\Theta_{2};\Theta_{a},\Theta_{3}\right),\left(\Theta_{U},\Theta_{1};\Theta_{\a},\Theta_{2};\Theta_{a},\Theta_{3}\right)\right] \,.
\ee
By generalizing the expression \eqref*{defMC2} to the case of centrally extended groups, the solution of this equation can be shown to be given by \eqref*{mcform}, plus the central terms
\be\eqlabel{mcformce}
\ba
\Theta_{1} \left( \dot U, \dot \a, \dot a \right)=  \left. \dfrac{d}{d\lambda}\Big(\mathcal{C}\left(U^{-1}(t),U\left( \lambda\right)\right)\Big)\right|_{\lambda=t} \,,\\[7pt]
\Theta_{2} \left( \dot U, \dot \a, \dot a \right)=\left\langle \cS\left(U\right),\Theta_{\a}\left( \dot U, \dot \a, \dot a \right) \right\rangle_\mfk \,,\\[7pt]
\Theta_{3} \left( \dot U, \dot \a, \dot a \right)=\left\langle \cS\left(U\right),\Theta_{a}\left( \dot U, \dot \a, \dot a \right) \right\rangle_\mfk +\frac{1}{2}\left\langle \mathfrak{s}\left(\alpha\right),\dot\alpha\right\rangle_\mfk \,.
\ea
\ee
Using the pairing \eqref*{pairingcen}, the action \eqref*{geomactionce} takes the form
\be\eqlabel{geomactionce2}
I_{\hat{H}}=I_{H}+c_1\int dt\, \Theta_1 \left( \dot U, \dot \a, \dot a \right)
+c_2 \int dt\, \Theta_2 \left( \dot U, \dot \a, \dot a \right)
+c_3\int dt\, \Theta_3 \left( \dot U, \dot \a, \dot a \right) \,,
\ee
where $I_H$ is the geometric action associated to the group $H$, given by expression \eqref*{espga2}.

\subsubsection*{Maxwell WZW model}
In the case of the $\hatLMax$ group \eqref*{hatLMax}, the Maurer-Cartan forms \eqref*{mcformce} can be evaluated using \eqref{mcform,comk,souriaulsl}. This leads to the following geometric action:
\be
I_{\hatLMax}= I_{\LMax}+I_{\rm WZW} \,,
\ee
where $I_{\LMax}$ is the action \eqref*{actionLMax} and $I_{\rm WZW}$ takes the form
\be\eqlabel{mwzw}
\ba
&&I_{\rm WZW} =  \dfrac{c_{1}}{4\pi} \displaystyle\int_0^{2\pi}  \hspace{-5pt} d\phi \int dt \, {\rm Tr} \left[U^{-1} U^{\prime}U^{-1} \dot U\right] +\dfrac{c_{1}}{4\pi}\int dt \, \int_{D^1}{\rm Tr}\left[\left( U^{-1} \bar{\rm d}U \right)^2 U^{-1} \dot U\right] \\[6pt]
&& + \dfrac{c_{2}}{2\pi} \displaystyle \int_0^{2\pi} \hspace{-5pt} d\phi \int  dt \,
\mathrm{Tr}\left[U^{-1}U^{\prime} \dot\a \right]  +\dfrac{c_{3}}{2\pi} \int_0^{2\pi}  \hspace{-5pt} d\phi \int dt {\rm Tr} \left[ U^{-1}U^{\prime}\left( \dot a-\frac{1}{2}\left[\a, \dot\a \right] \right) +\frac{1}{2}\alpha^{\prime} \dot\a\right] \,.
\ea
\ee
Here the exterior derivative $\bar{\rm d}$ is defined on $D^1$, as in \eqref*{comk}. This corresponds to a chiral WZW model, and it is globally invariant under the $\hatLMax$ group. The first term in \eqref{mwzw} corresponds to the usual chiral WZW model that defines the geometric action associated to the $\hatLsl$ Kac-Moody group \cite{Alekseev:1988ce,Aratyn:1990dj,Alekseev:2018ful}. The second term corresponds to the flat chiral WZW model, which has been studied in \cite{Salomonson:1989fw,Barnich:2013yka,Barnich:2017jgw} in the context of three-dimensional gravity\footnote{In order to make this term match the model found in \cite{Salomonson:1989fw,Barnich:2013yka,Barnich:2017jgw}, one has to replace the pair $(U,\a)$ by its inverse $(U,\a)^{-1}=(U^{-1},-\Ad_{U^{-1}}\a)$, and integrate by parts.}. The third term is a novel chiral WZW model by itself and defines a Maxwell extension of the known WZW models previously discussed.

\subsubsection*{Maxwell gravitational WZ model}

Let us consider again the $\MhatBMS$ group defined in \eqref*{MhatBMS}. In this case the Maurer-Cartan forms \eqref*{mcformce}, and the geometric action \eqref*{geomactionce2} can be evaluated using  \eqref*{mcform,btcc,souriauvir}, which yields
\be
I_{\MhatBMS}= I_{\MBMS}+I_{\rm gWZ} \,,
\ee
where $I_{\MBMS}$ corresponds to the geometric action associated to the centerless $\MBMS$ group, given by \eqref*{actionMBMS}, and $I_{\rm gWZ}$ reads
\be\eqlabel{gwz}
\ba
I_{\rm gWZ}= \displaystyle \frac{c_1}{48\pi}\int_0^{2\pi}  \hspace{-5pt} d\phi \int dt\,  \left(\frac{U^{\prime\prime}}{U^\prime} \right) ^{\prime} \frac{\dot U}{U^\prime}+ \frac{c_2}{24\pi}\int_0^{2\pi}  \hspace{-5pt} d\phi \int  dt \, \left(\frac{\a\circ U}{U^\prime}
\right)^{\prime\prime\prime} \frac{\dot U}{U^\prime} \\[6pt]
\displaystyle + \frac{c_3}{24\pi} \int_0^{2\pi}  \hspace{-5pt} d\phi \int dt \, \left(\frac{a\circ U}{U^\prime} \right)^{\prime\prime\prime} \frac{\dot U}{U^\prime}+ \frac{c_3}{48\pi} \int_0^{2\pi}  \hspace{-5pt} d\phi \int  dt \, \left(\frac{\a\circ U}{U^\prime}\right)^{\prime\prime\prime} \frac{\dot \a\circ U}{U^\prime} \,.
\ea
\ee
The action $I_{\rm gWZ}$ defines a Maxwell extension of the gravitational Wess-Zumino action \cite{Alekseev:1988ce,Aratyn:1990dj,Kubo:1994cf,Alekseev:2018ful}, found as the geometric action describing world lines on the Virasoro group. In fact, by defining the change of variables
\be 
F=U^{-1}, \quad \varphi=U(\phi) \,,
\ee
the first term in \eqref*{gwz} (supplemented with the first term in \eqref*{actionMBMS}) takes exactly the same form as the Polyakov action found in \cite{Polyakov:1987zb} in the context of induced two-dimensional quantum gravity:
\be
\int_0^{2\pi}  \hspace{-5pt} d\phi \int dt \,\frac{\dot U}{U^\prime} \left[ J_0 + \frac{c_1}{48\pi} \left(\frac{U^{\prime\prime}}{U^\prime} \right) ^{\prime}  \right]
 \rightarrow
\int_0^{2\pi} \hspace{-5pt}  d\varphi \int  dt \,  \dot F \left[   F^\prime J_0 (F)  -\frac{c_1}{48\pi}   \left(   \frac{F^{\prime\prime\prime}}{F^{\prime\,2}}- 2 \frac{F^{\prime\prime\,2}}{F^{\prime\,3}}   \right)    \right] \,.
\ee
The second term in \eqref*{gwz} corresponds to the $\BMS$ extension of the gravitational WZ action, found in \cite{Barnich:2017jgw} as the geometric action on the $\BMS$ group. It is related to the flat limit of Liouville theory, which defines  the classical dual of asymptotically flat three-dimensional Einstein gravity \cite{Barnich:2013yka}. The third term in \eqref*{gwz} defines a novel Extension of the Polyakov action for two-dimensional gravity, which is by definition invariant under the $\MBMS$ group.

\section{Discussion and possible applications}
\label{sec:discussion}

In this paper, we have analysed the Maxwell group in 2+1 dimensions and its infinite-dimensional generalizations. This have been made by extending the notion a semi-direct product group in a way that captures the underlying structure of the Maxwell symmetry. The definition of extended semi-direct product groups allowed us to define the adjoint and the coadjoint representation of Maxwell groups in a general fashion, and to construct geometric actions on their coadjoint orbits . Subsequently, we have explicitly shown how the $\Maxd$ group fits in this description. The geometric action in this case leads to the description of a relativistic particle coupled to a constant electromagnetic field. Subsequently, we have considered the semi-direct product of a Lie group and a central extension of its Lie algebra. When based on the $\Sl$ symmetry, this group structure matches the double cover of the Maxwell group in 2+1 dimensions. Similarly, higher-spin extensions of the Maxwell group can be constructed by considering these special kind of extended semi-direct products. 

Special extended semi-direct products admit central extensions, for which a general description has been given. Using this construction, two infinite-dimensional enhancements of the Maxwell symmetry have been constructed, namely, the $\hatLMax$ Kac-Moody group and the $\MhatBMS$ group. These symmetries define Maxwell-like extensions of the ${\rm ISL}\left(2,\mathbb{R}\right)$ Kac-Moody group and the $\hatBMS$ group, respectively. The corresponding Lie algebras of these infinite-dimensional groups have been previously studied in the context of algebra expansions and three-dimensional gravity \cite{Caroca:2017onr,Concha:2018zeb}. The geometric actions associated to these infinite-dimensional groups lead to novel Wess-Zumino terms that generalize the known results presented, for instance, in \cite{Alekseev:2018ful,Barnich:2017jgw}. The first geometric action is a novel WZW model, invariant under the Maxwell-Kac-Moody group $\hatLMax$. The second case that we considered is the geometric action associated to the $\MhatBMS$ group, which defines a Maxwell extension of gravitational Wess-Zumino action first found by Polyakov as an action for two-dimensional gravity \cite{Polyakov:1987zb}. In the following, we briefly elaborate on the application of these geometric actions and comment on possible future directions.

\subsubsection*{Chern-Simons theory and quantum Hall systems}

A (2+1)-dimensional field theory that naturally realizes the $\hatLMax$ Kac-Moody symmetry at the level of its conserved charges is a Chern-Simons theory invariant under the $\max$ algebra \eqref*{maxalg2+1}:
\begin{equation*}
I_{\rm CS}=\frac{k}{4\pi} \int_M \left\langle A\wedge{\rm d}A+\frac{2}{3} A\wedge A\wedge A \right\rangle \,,
\end{equation*}
where $A$ is a $\max$-valued connection one-form on a manifold $M$, and $\left\langle\;\;\right\rangle$ denotes an invariant metric on $\max$. The conserved charges in a Chern-Simons theory are known to satisfy a Kac-Moody algebra that enhances the original symmetry of the action \cite{Banados:1994tn}. It is therefore evident that, as in this case the gauge group is given by $\Max$, the conserved charges of the theory will satisfy a Poisson algebra isomorphic to the Maxwell-Kac-Moody algebra \eqref*{maxkmalg}. Hence, the $\hatLMax$ Kac-Moody group can be relevant in any context involving a $\Max$-invariant Chern-Simons action.

Chern-Simons forms provide effective theories for the large scale dynamics of quantum Hall systems \cite{Zhang:1988wy,Bahcall:1991an,Fradkin:1991wy,Balachandran:1991nw} (for a recent review see \cite{Tong:2016kpv}). In\cite{Palumbo:2016nku}, a Chern-Simons action invariant under the $\max$ algebra was used to construct an effective theory for a topological insulator. This leads to a relativistic version of the Wen-Zee term, which couples the background geometry to the Chern-Simons effective field in quantum Hall fluids \cite{Wen:1992ej}. The motivation to use the Maxwell symmetry in this description stems from the fact that it includes non-commuting translations, allowing one to reproduce the Girvin-MacDonald-Plazmann algebra of magnetic translations \cite{girvin1985collective}. As stated before, the current algebra of such Chern-Simons theory is naturally given by the Maxwell-Kac-Moody algebra $\hatLmax$. This symmetry can be made explicit at the level of the action when studying its edge modes. In fact, the boundary dynamics of a Chern-Simons theory is described by WZW model, whose kinetic term has the general form
\begin{equation*}
I_{\rm WZW}=\frac{k}{4\pi}\int_0^{2\pi} \hspace{-5pt} d\phi \int dt \, \left\langle g^{-1}\dot gg^{-1} g^\prime \right\rangle -\frac{k}{12\pi}\int_M\left\langle\left(g^{-1}{\rm d}g\right)^3\right\rangle \,,
\end{equation*}
where $t$ and $\phi$ are coordinates on the boundary $\partial M$. It is easy to show that, considering and element of the $\Max$ group \eqref*{dmaxwell2+1} in the form $g=(U,\a,a)^{-1}$, and the invariant metric on $\max$
\begin{equation*}
\left\langle \bJ_\mu \bJ_\nu \right\rangle= \gamma_1 \eta_{\mu\nu}  \,,\quad
\left\langle \bJ_\mu \bPi_\nu \right\rangle= \gamma_2 \eta_{\mu\nu}  \,,\quad
\left\langle \bPi_\mu \bPi_\nu\right\rangle= \gamma_3 \eta_{\mu\nu}=\left\langle \bJ_\mu \bF_\nu \right\rangle \,,
\end{equation*}
this action is exactly the WZW model \eqref*{mwzw}, obtained in Section \ref{wzwterms} (where the Chern-Simons level $k$ is related to the central charges through the constants entering in the invariant metric on the $\max$ algebra). Thus, the geometric
action \eqref*{mwzw} describes the boundary dynamics of a $\Max$-invariant Chern-Simons theory, and naturally provides an action for the edge modes of the topological insulator model introduced in \cite{Palumbo:2016nku}. One way to describe edge modes in such a system is to consider a dislocation at the boundary of a topological insulator in 3+1 dimensions as considered, for instance, in \cite{Teo:2010zb}. 

\subsubsection*{Three-dimensional gravity}

Chern-Simons theories provide a gauge theoretical formulation for gravity in 2+1 dimensions \cite{Witten:1988hc}. Along these lines, a Chern-Simons theory invariant under the Maxwell algebra \eqref*{maxalg2+1} has been studied as an extension of three-dimensional Einstein gravity \cite{Salgado:2014jka,Hoseinzadeh:2014bla}. In \cite{Concha:2018zeb}, it was shown that the asymptotic symmetry of such a gravity theory is given by an extension of the $\bms$ algebra, given precisely by the $\mhatbms$ algebra \eqref*{mbms3}.  This is a generalization of the known results for asymptotically AdS and flat gravity in 2+1 dimensions, where the asymptotic symmetry group is known to be given by the conformal algebra in two dimensions and the $\BMS$ group, respectively \cite{Brown:1986nw,Barnich:2006av}. When the role of the generators of the extended space of translations \eqref*{idvext} are interchanged, this leads to an extended gravity theory invariant under the simplest example of a Hietarinta algebra \cite{Bansal:2018qyz}. This algebra is isomorphic to a particular conformal extension of the Galilean algebra, and its asymptotic symmetry is also given by the extended semidirect sum \eqref*{mbms3algebra}, where the  generators $\bPP_m$ and $\bFF_m$ in \eqref{basembms3} are interchanged \cite{Chernyavsky:2019hyp}. In this article, we have succeeded in finding a group structure that leads to this infinite-dimensional algebra, namely, the $\MhatBMS$ group \eqref*{MBMS}. The relation of this group with Chern-Simons gravity can be analysed in the following way. The solution for the gauge connection used in \cite{Concha:2018zeb} is given, up to a gauge transformation depending on the radial coordinate, by
\begin{equation*}
\begin{aligned}
A &=  \frac{1}{2}\Big({\cal M}dt+\left({\cal J}+t{\cal M}^\prime\right)d\phi\Big)\,\bPi_{0}+dt\,\bPi_{1}\\
&+\frac{1}{2}{\cal \,M}d\phi\,\bJ_{0}+d\phi\,\bJ_{1}+\frac{1}{2}\Big(\left({\cal J}+t{\cal M}^\prime\right)du+\left({\cal Z}+t{\cal J}^\prime+\frac{t^2}{2}{\cal M}\prime\prime\right)d\phi\Big)\,\bF_{0}\,,
\end{aligned}
\end{equation*}
and depends only on the arbitrary functions of the boundary coordinates, given by $\mathcal{M}$, $\mathcal{J}$ and $\mathcal{Z}$. Under the action of an improper gauge transformation, they change as
\begin{equation*}
\ba
\delta{\cal M} & = & {\cal M}^{\prime}Y+2{\cal M}Y^{\prime}-2Y{}^{\prime\prime\prime}\,, \\[5pt]
\delta{\cal J} & = & {\cal M}^{\prime}T+2{\cal M}T^{\prime}-2T{}^{\prime\prime\prime}+{\cal J}^{\prime}Y+2{\cal J}Y^{\prime}\,,\\[5pt]
\delta{\cal Z} & = & {\cal M}^{\prime}R+2{\cal M}R^{\prime}-2R{}^{\prime\prime\prime}+{\cal J}^{\prime}T+2{\cal J}T^{\prime}+{\cal Z}^{\prime}Y+2{\cal Z}Y^{\prime}\,,
\ea
\end{equation*}
which defines the solution space of the theory. It is a straightforward computation to show that this expression matches the infinitesimal form of coadjoint action $\Ad_{\left( Y,m_1; T,m_2 ; R,m_3  \right)} \left( \mathcal{M}, c_1 ; \mathcal{N}, c_2; \mathcal{F},c_3 \right)$ given in  \eqref*{Adcentrallyext0}, when the Souriau cocycle on the Virasoro algebra \eqref*{souriauvir}, and the paring \eqref*{pairingwitt} are implemented. This shows that, when adopting the boundary conditions introduced in \cite{Concha:2018zeb} , the solution space of the $\Max$-invariant Chern-Simons theory matches the coadjoint representation of the $\MhatBMS$ group presented here. On the other hand, a coadjoint orbit of the $\MhatBMS$ group with representative $(J_0,\Pi_0,F_0)$ defines a phase space for the geometric action \eqref*{gwz}. Therefore, the Maxwell extension of the gravitational WZ action can be conjectured to control the solution space of this Chern-Simons theory, when the boundary conditions found in \cite{Concha:2018zeb} are adopted. This is an interesting subject for future research.

On the other hand, chiral WZW models appear in three-dimensional gravity as an intermediate
point in the Hamiltonian reduction that leads to the corresponding
classical dual defined at the boundary \cite{Coussaert:1995zp}. In our case, the Maxwell WZW model plays this role. This suggest there should be way to connect the Maxwell WZW model \eqref*{mwzw} with the gravitational Maxwell WZ action \eqref*{gwz} by a Drinfeld-Sokolov reduction \cite{Bershadsky:1989mf}. This is a work in progress.

Kac-Moody algebras can also provide asymptotic symmetries for gravity theories when the most general boundary conditions are adopted \cite{Grumiller:2016pqb}. In this context, allowing the most general boundary conditions for the $\max$ Chern-Simons connection, would lead to the Maxwell-Kac-Moody algebra $\hatLmax$ as the asymptotic symmetry.

\subsubsection*{Higher-spin symmetries}
In Section \ref{maxalgebra2+1}, we have shown how to construct higher-spin extensions of the $\Max$ group, which are given by extended semi-direct products of the form ${\rm SL}\left(N,\mathbb{R}\right) \ltimes_\Ad \sln_{\rm ext}^{\rm (ab)}$. In the same way, higher-spin extensions of the $\mhatbms$ algebra  can be constructed defining extended semi-direct sums based on the $\mathcal{W}_N$ algebra.
\begin{equation*}
\mathcal{W}_N \oright_\ad \left( \mathcal{W}_N\right)^{\rm (ab)}_{\rm ext} \,.
\end{equation*}
This is interesting in the context of gravity, where $\mathcal{W}_N$ algebras are known to describe the asymptotic dynamics of higher-spin extensions of Einstein gravity in 2+1 dimensions \cite{Campoleoni:2010zq,Perez:2014pya,Banados:2015tft}. In our case, these algebras define Maxwell extensions of the asymptotic symmetries found in flat higher-spin gravity \cite{Gonzalez:2013oaa,Matulich:2014hea,Ammon:2017vwt}, given by semi-direct sums based on the $\mathcal{W}_N$ algebra \cite{Campoleoni:2016vsh}. Thus, infinite-dimensional algebras of this kind are likely to define asymptotic symmetries of higher-spin gravity theories based on the $\Max$ group, like the one presented in \cite{Caroca:2017izc}. On the other hand, $\mathcal{W}_N$ algebras have also been found in the description of quantum Hall systems \cite{Iso:1992aa,Cappelli:1992yv}. Based on the fact that a $\Max$-invariant Chern-Simons theory captures features of the magnetic translations at the boundary of a topological insulator, it would be interesting to evaluate if Maxwell extensions of the $\mathcal{W}_N$ symmetry could somehow emerge in this type of systems.

\section*{Acknowledgements}
%\eqlabel{sec:acknowledgements}

\addcontentsline{toc}{section}{Acknowledgments}

The author would like to thank P. Brzykcy, N. Merino, G. Palumbo, D. Sorokin and J. Tosiek for useful suggestions and comments. Special acknowledgements are given to J. Gomis for valuable discussions and detailed explanations about different aspects of the Maxwell symmetry, and to B. Mudza for a careful reading of the manuscript. P. Salgado-Rebolledo acknowledges DI-VRIEA for financial support through Proyecto Postdoctorado 2019 VRIEA-PUCV. Partial support by the Chilean FONDECYT grant N$^\circ$3160581 is also acknowledged. 

\providecommand{\href}[2]{#2}\begingroup\raggedright\endgroup

\end{document}